\documentclass[11pt, a4paper]{article}

\usepackage[dvips,letterpaper]{geometry}
\usepackage{graphicx}
\usepackage{times}
\usepackage{setspace}
\usepackage{amsthm, amsmath, natbib, amssymb, amsfonts}
\usepackage{mathtools, here}
\usepackage{breakcites}
\usepackage{bm}
\usepackage{multirow}
\usepackage{subfigure}
\usepackage{url}
\usepackage{verbatim}
\usepackage{blkarray}
\usepackage{natbib}
\usepackage{color}
\usepackage{booktabs}
\usepackage{physics}
\usepackage{lineno}
\usepackage{listings}
\usepackage{tikz}
\usetikzlibrary{matrix,shapes,arrows,positioning}
\definecolor{codegreen}{rgb}{0,0.6,0}
\definecolor{codegray}{rgb}{0.5,0.5,0.5}
\definecolor{codepurple}{rgb}{0.58,0,0.82}
\definecolor{backcolour}{rgb}{0.95,0.95,0.92}

\lstdefinestyle{mystyle}{
    backgroundcolor=\color{backcolour},
    commentstyle=\color{codegreen},
    keywordstyle=\color{magenta},
    numberstyle=\tiny\color{codegray},
    stringstyle=\color{codepurple},
    basicstyle=\footnotesize,
    breakatwhitespace=false,
    breaklines=true,
    captionpos=b,
    keepspaces=true,
    numbers=left,
    numbersep=5pt,
    showspaces=false,
    showstringspaces=false,
    showtabs=true,
    tabsize=2}

\input cyracc.def

\usepackage[breaklinks=true,pdfencoding=auto]{hyperref}

\makeatletter
\newcommand\fake@math{}% just for safety
\def\fake@math#1\){[math]}
\makeatother

\def \build#1#2#3{\mathrel{\mathop{#1}\limits^{#2}_{#3}}}

\renewcommand{\Hat}{\widehat}

\newtheorem{Remark}{Remark}[section]

\title{A new regression model for positive data}
 \vspace{-0.25cm}
\author{\normalsize
\textbf{Marcelo Bourguignon}$^{1}$\,\, \textbf{Manoel Santos-Neto}$^{2,3}$ \,\,\textbf{and} \,\,  \textbf{M\'ario de Castro}$^{4}$
\\
{\footnotesize $^{1}$Departamento de Estat\'istica, Universidade Federal do Rio Grande do Norte, Natal, Brazil}\\[-0.15cm]
{\footnotesize $^{2}$Departamento de Estat\'istica, Universidade Federal de S\~ao Carlos, S\~ao Carlos, Brasil}\\[-0.15cm]
{\footnotesize $^{3}$Departamento de Estat\'istica, Universidade Federal de Campina Grande, Campina Grande, Brasil}\\[-0.15cm]
{\footnotesize $^{4}$Instituto de Ci\^encias Matem\'aticas e de Computa\c{c}\~ao, Universidade de S\~ao Paulo, S\~ao Carlos, Brasil}\\[-0.15cm]
}

\date{}

\begin{document}
%\linenumbers

\maketitle

\vspace{-0.7cm}
\begin{abstract}
In this paper, we propose a regression model where the response variable is beta prime distributed using a new parameterization of this distribution that is indexed by mean and precision parameters. The proposed regression model is useful for situations where the variable of interest is continuous and restricted to the positive real line and is related to other variables through the mean and precision parameters.  The variance function of the proposed model has a quadratic form. In addition, the beta prime model has properties that its competitor distributions of the exponential family do not have. Estimation is performed by maximum likelihood. Furthermore, we discuss residuals and influence diagnostic tools. Finally, we also carry out an application to real data that demonstrates the usefulness of the proposed model.\\

\vspace{-0.3cm}

%\vspace{-0.3cm}
%\paragraph{Mathematics subject classification} Primary 62F10; Secondary 62F12.
\end{abstract}

\section{Introduction}\label{sec:1}

The concept of regression is very important in statistical data analysis~\citep{bent:97}.
In this context, generalized linear models~\citep{Nelder:1972aa}
are regression models for response variables in the exponential family.
Let $Y_1, \ldots, Y_n$ be $n$ independent random variables. According to~\cite{McNeld:89}, the random component of a
generalized linear model (GLM) is specified by the probability density function
$
f(y| \theta, \phi) = \exp\{\phi[y\,\theta - b(\theta)] + c(y, \phi)\},
$
where $\theta$ is the canonical parameter, $b(\cdot)$ and $c(\cdot, \cdot)$ are known appropriate functions. The mean and the variance
of $Y_i$ are
\begin{equation*}\label{momentsMLG}
\textrm{E}[Y_i] = \mu_i = \dv{b(\theta_i)}{\theta_i} \quad \textrm{and} \quad \textrm{Var}[Y_i] = \frac{\textrm{V}(\mu_i)}{\phi},
\end{equation*}
respectively, where $\textrm{V}(\mu_i) = \dv{\mu_i}{\theta_i}$ is called the variance function, $\phi > 0$ is regarded
as a precision parameter and $\mu$ vary in an interval of the real line.
The variance function characterizes the distribution. For the normal, gamma, inverse Gaussian, binomial
and Poisson distributions, we have $\textrm{V}(\mu_i) = 1, \textrm{V}(\mu_i) = \mu_i^2, \textrm{V}(\mu_i) = \mu_i^3, \textrm{V}(\mu_i) = \mu_i(1-\mu_i)$ and $\textrm{V}(\mu_i) = \mu_i$,
respectively. However, the linear exponential families with given mean-variance relationships do not always exist~\citep{Bar-Lev:1986aa}.

The main aim of this paper is to propose a regression model that is tailored for situations where the response variable is measured continuously on the  positive real line that is in several aspects, like the generalized linear models.
In particular, the proposed model is based on the assumption that the response is beta prime (BP) distributed. We considered a new parameterization of the BP distribution in terms of the mean and precision parameters. Under this parameterization, we propose a  regression model, and we allow a regression structure for the mean and precision parameters by considering the mean and precision structure separately. The variance function of the proposed model assumes a quadratic form. The proposed regression model is convenient for modeling asymmetric data, and it is an alternative to the generalized linear models when the data presents skewness (see Section 2). Inference, diagnostic and selection tools for the proposed class of models will be presented.

The BP distribution~\citep{keep:62,McDonald:1984aa} is also known as inverted beta distribution or beta distribution of the second kind.
However, only a few works have studied the BP distribution. \cite{McDonald:1987aa} discussed its properties and obtained the maximum likelihood estimates of the model parameters. \cite{McDonald:1990aa} and \cite{Tulupyev:2013aa} used the BP distribution while discussing regression models for positive random variables. However, these works have considered the usual parameterization of the BP distribution. For more detail and some generalizations of BP distribution, see \cite{johkotbala:95}.

A random variable $Y$ follows the \text{BP} distribution
with shape parameters $\alpha > 0$ and $\beta > 0$, denoted by
$Y \sim \text{BP}(\alpha,\beta)$, if its cumulative distribution function
(cdf) is given by
\begin{equation}\label{int:01}
F(y| \alpha, \beta) = I_{y/(1+y)}(\alpha, \beta),
 \quad {y}>0,
\end{equation}
where $I_{x}(\alpha, \beta) = B_x(\alpha, \beta)/B(\alpha, \beta)$ is the incomplete beta function ratio,
$B_x(\alpha, \beta) = \int_{0}^{x}\omega^{\alpha-1}(1-\omega)^{\beta-1}\textrm{d} \omega$ is the incomplete function,
$B(\alpha, \beta) = \Gamma(\alpha)\Gamma(\beta)/\Gamma(\alpha + \beta)$ is the beta function and
$\Gamma(\alpha) = \int_{0}^{\infty}\omega^{\alpha-1}\textrm{e}^{-\omega}\textrm{d} \omega$ is the gamma function.

The probability density function (pdf) associated with \eqref{int:01} is
\begin{equation}\label{inv:01}
f(y| \alpha, \beta) = \frac{y^{\alpha-1}(1 + y)^{-(\alpha+\beta)}}{B(\alpha, \beta)}, \quad {y}>0.
\end{equation}

%Also, the hazard function (also known as failure rate) for the beta prime model is given by
%\begin{equation}
%h(y| \alpha, \beta) = \frac{f(y| \alpha, \beta)}{1-F(y| \alpha, \beta)} = \frac{y^{\alpha-1}(1 + y)^{-(\alpha+\beta)}}{B(\alpha, \beta)- B_{y/(1+y)}(\alpha, %\beta)}.
%\label{eq:hfcbp}
%\end{equation}

It can be proved that the BP
density function is decreasing with $f(y| \alpha, \beta) \rightarrow \infty$ as $y \rightarrow 0$, if $0 < \alpha < 1$;
$f(y| \alpha, \beta)$ is decreasing with mode at $y = 0$, if $\alpha = 1$; and $f(y| \alpha, \beta)$ increases and then decreases with mode at
$y = (\alpha - 1)/(\beta + 1)$ for $\alpha > 1$. Furthermore, if $0 < \alpha \leq 1$, $f(y| \alpha, \beta)$ is concave; if $1 < \alpha \leq 2$,
$f(y| \alpha, \beta)$ is convex and then upward, with inflection point at $x_2$; and if $\alpha > 2$, $f(y| \alpha, \beta)$ is concave, then downward, then upward again, with inflection points at $x_1$ and $x_2$, where
$$x_1 = \frac{(\alpha-1)(\alpha+2) - \sqrt{(\alpha-1)(\beta+2)(\alpha+\beta)}}{(\beta+2)(\beta+1)}$$
and
$$x_2 = \frac{(\alpha-1)(\alpha+2) + \sqrt{(\alpha-1)(\beta+2)(\alpha+\beta)}}{(\beta+2)(\beta+1)}.$$

There are some stochastic representation of the BP random variable. Let $Z$ follow the beta distribution with positive parameters $\alpha$ and $\beta$, then
\begin{equation}\label{re1}
Y \stackrel{d}{=} \frac{Z}{1-Z},
\end{equation}
where $\stackrel{d}{=}$ stands for equality in distribution. Furthermore, let $V$ and $W$ be independent random variables following standard
(scale parameter equal to 1) gamma distributions with shape parameters $\alpha > 0$ and $\beta > 0$, respectively, i.e.,
$V \sim \text{Ga}(\alpha, 1)$ and $W \sim \text{Ga}(\beta, 1)$, then
\begin{equation}\label{re2}
Y \stackrel{d}{=} \frac{V}{W}.
\end{equation}
Note that we can use the stochastic representations (\ref{re1}) and (\ref{re2}) for estimating the parameters of the BP model (based on the EM-algorithm).

\begin{Remark}\label{re01}
The BP distribution can be written in the exponential family form, but is not a dispersion model. Then, the BP distribution has a two-parameter general exponential distribution with natural parameters $\alpha-1$ and $-(\alpha+\beta)$ and natural sufficient statistics $\log(Y)$ and $\log(1+Y)$.
\end{Remark}

The $r$th moment about zero of $Y$ is given by
\begin{equation*}\label{inv:04}
\textrm{E}[Y^r] = \frac{B(\alpha + r, \beta - r)}{B(\alpha, \beta)}, \quad -\alpha < r < \beta.
\end{equation*}
For $r \in \mathbb{N}$ and $r < \beta$, the $r$th moment about zero simplifies to
\begin{equation*}\label{inv:044}
\textrm{E}[Y^r] = \prod\limits_{i=1}^{r}\frac{\alpha + i - 1}{\beta-i}.
\end{equation*}
In particular, the mean and the variance associated with \eqref{inv:01} are given by
\begin{equation}\label{inv:05}
\textrm{E}[Y] = \frac{\alpha}{\beta-1}, \quad \beta > 1, \quad
\textrm{and} \quad
\textrm{Var}[Y]= \frac{\alpha(\alpha + \beta - 1)}{(\beta-2)(\beta-1)^2}, \quad \beta > 2,
\end{equation}
respectively. Furthermore, $\textrm{E}[\log (Y)] = \Psi^{(0)}(\alpha) - \Psi^{(0)}(\beta)$ and $\textrm{E}[\log (1 + Y)] = \Psi^{(0)}(\alpha + \beta) - \Psi^{(0)}(\beta)$,
where $\Psi^{(0)}(z) = \dv{\ln[\Gamma(z)]}{z}$ is the digamma function.

Some other properties of the BP distribution are as follows. If $Y \sim \text{BP}(\alpha,\beta)$, then: (i) $1/Y \sim \text{BP}(\beta,\alpha)$, that is, the BP distribution is closed under
reciprocation; (ii) $\frac{\beta}{\alpha} Y \sim \textrm{F}(2\, \alpha, 2\,\beta)$, that is, the BP distribution is closed under scale transformations;
(iii) $Y$ has positive skewness, but due to its flexibility, symmetric data can also be modeled by the BP distribution (see Figure \ref{fig1}); (iv) $Y$ has several stochastic representations, see (\ref{re1}) and (\ref{re2}). It should be highlighted that, all of these properties are shared by few distributions, in particular, several of them are not shared by exponential family distributions used in GLM.

%\begin{Proposition} Let $Y$ be a random variable with pdf given by (\ref{inv:01}). Then,\\
%\begin{enumerate}
%\item If $Y \sim \text{BP}(\alpha,\beta)$, then $1/Y \sim \text{BP}(\beta,\alpha)$ (The beta prime family is closed under the reciprocal transformation);
%\item If $Y \sim \text{BP}(\alpha,\beta)$, then $X = \frac{\beta}{\alpha} X \sim F(2\alpha,2\beta)$ (Connection with the F distribution);
%\item The standard BP distribution is the same as the standard log-logistic distribution.
%\end{enumerate}
%\end{Proposition}

Furthermore, the BP distribution has several related distributions. In particular,
\begin{itemize}
\item[i)] If $X$ has a Lomax distribution \citep{Lomax54}, also known as a Pareto Type II distribution, with shape parameter $\alpha$ and scale parameter $\lambda$, then $X/\lambda \sim \textrm{BP}(1, \alpha)$;

\item[ii)] If $X$ has a Pareto distribution \citep{Arnold} with minimum $x_{m}$ and shape parameter $\alpha$, then $X-x_{m} \sim \textrm{BP}(1, \alpha)$;

\item[iii)] If $X$ has a standard Pareto Type IV distribution \citep{Rodriguez77} with shape parameter $\alpha$ and inequality parameter $\gamma$, then $X^{\frac{1}{\gamma}} \sim \textrm{BP}(1, \alpha)$;

\item[iv)] If $X \sim \textrm{BP}(\alpha, \beta)$, then $\frac{\beta}{\alpha}X$ has the F distribution \citep{Phillips82} with $2\,\alpha$ degrees of the freedom in the numerator and $2\,\beta$ degrees of freedom in the denominator.
\end{itemize}

The rest of the paper proceeds as follows. In Section~\ref{sec:3}, we introduce a new parameterization of the BP distribution that is indexed by the mean and precision parameters.
Section~\ref{sec:4} presents the BP regression model with varying mean and precision. Diagnostic measures are discussed in Section~\ref{sec:5}. In Section ~\ref{sec:6}, some numerical results of the estimators are presented with a discussion of the obtained results. In Section~\ref{sec:7}, we discuss an application to real data that demonstrates the usefulness of the proposed model. Concluding remarks are given in the final section.

\section{A BP distribution parameterized by its mean and precision parameters}
\label{sec:3}

\begin{figure}[!htbp]
\centering
\subfigure[$\phi = 1$]{\includegraphics[height=5.5cm,width=5.5cm]{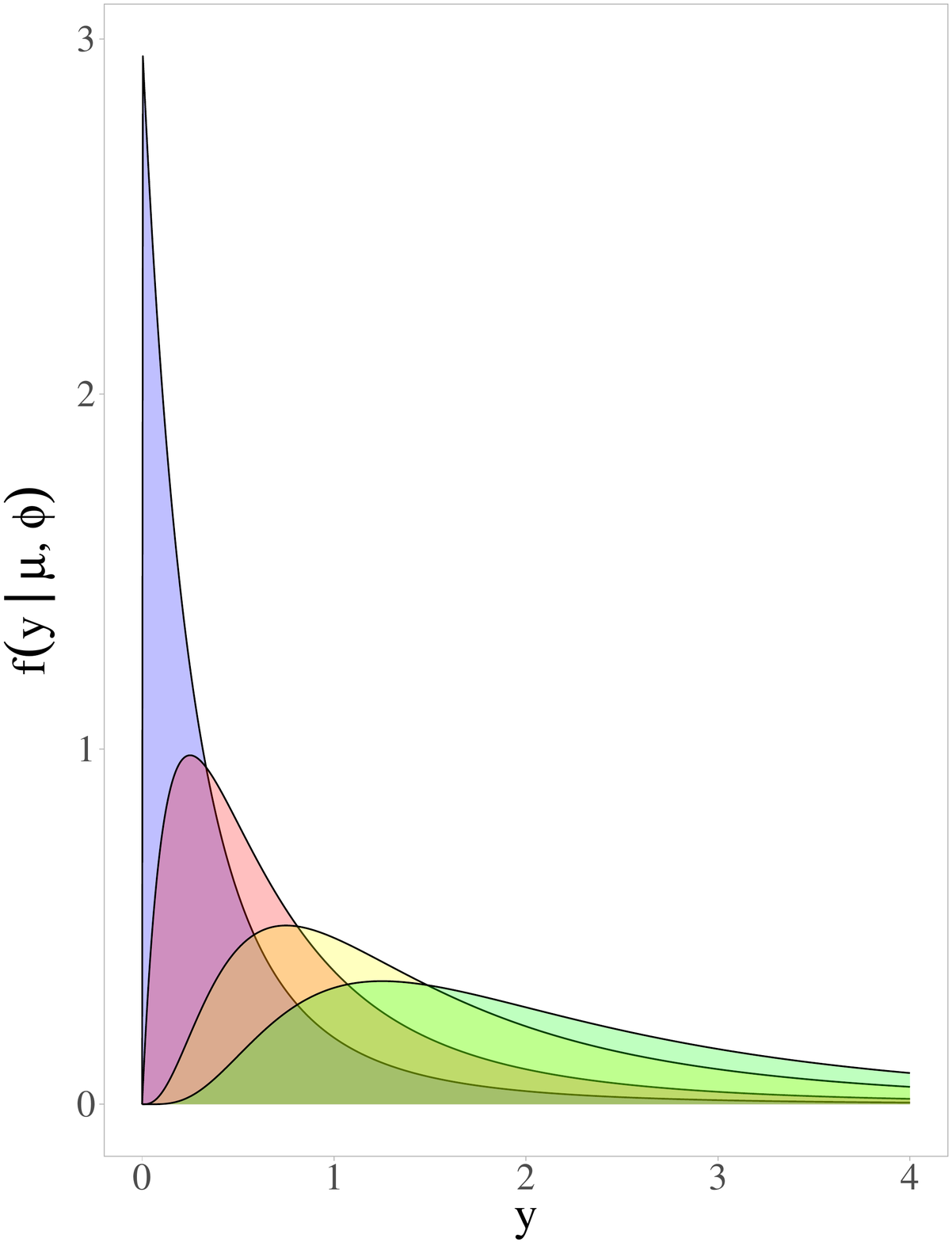}}
\subfigure[$\phi = 10$]{\includegraphics[height=5.5cm,width=5.5cm]{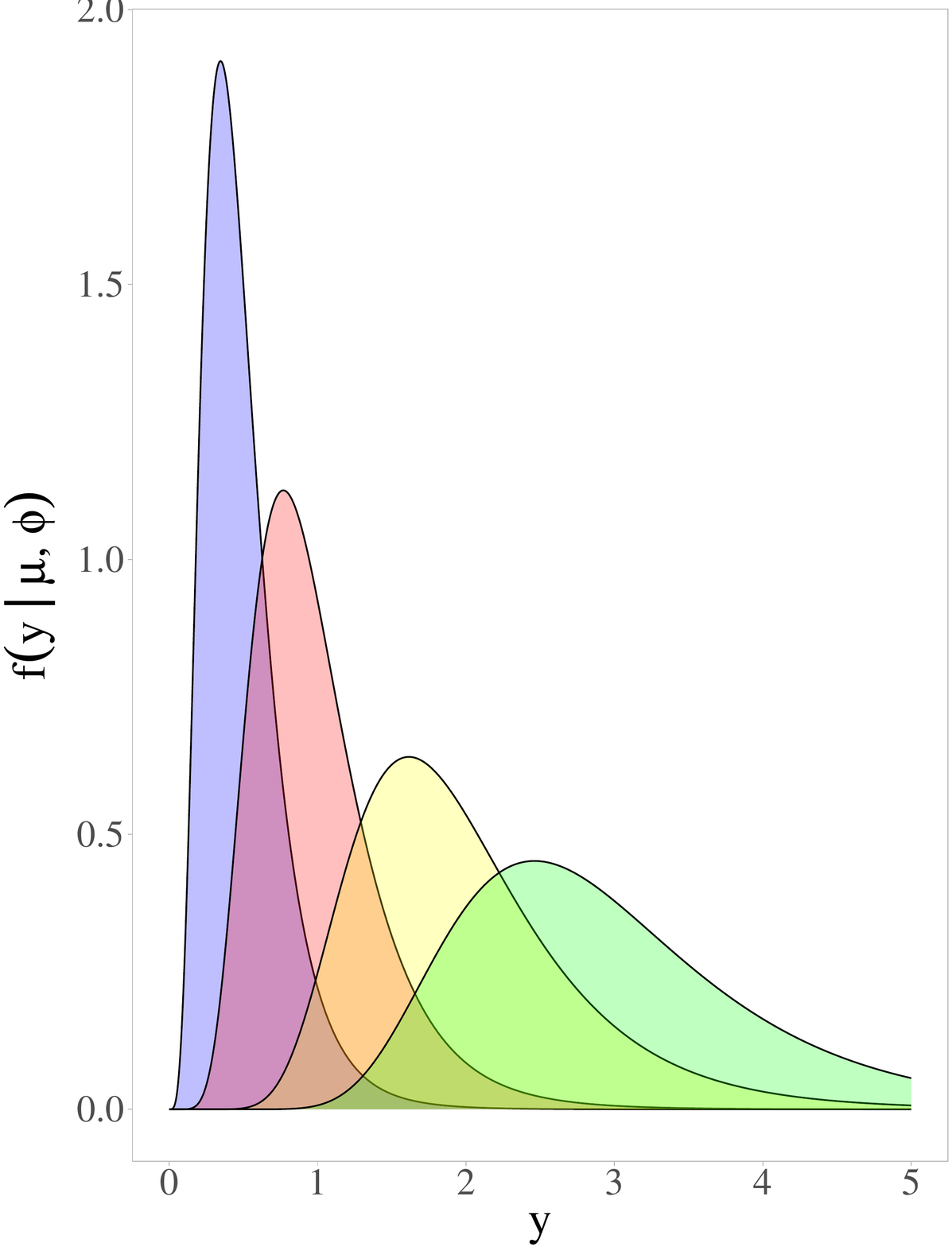}}
\subfigure[$\phi = 100$]{\includegraphics[height=5.5cm,width=5.5cm]{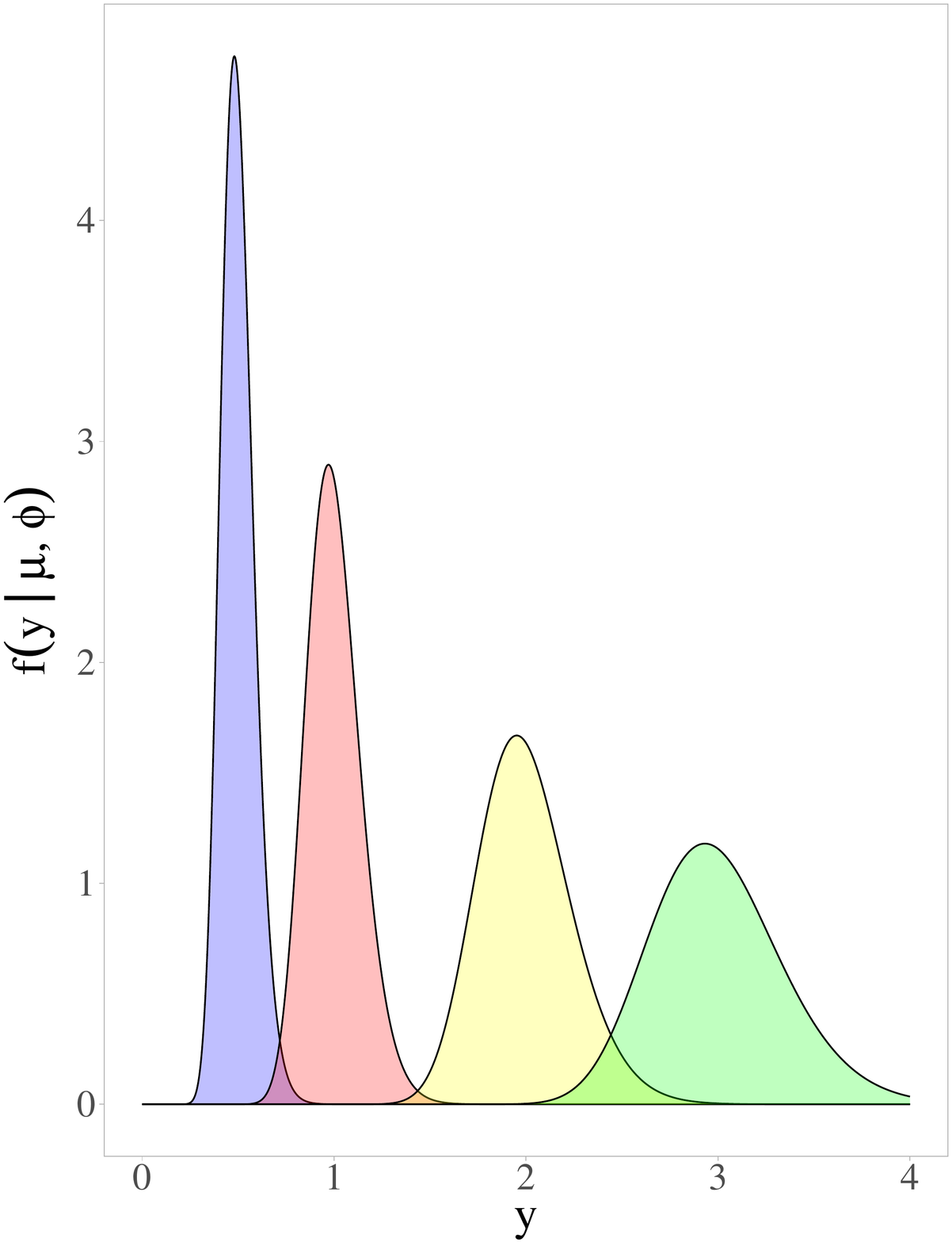}}
\caption{Plots of probability density function of the BP distribution considering the following values of $\mu$: $0.5$ (blue), $1.0$ (red), $2.0$ (yellow) and $3.0$ (green).}
\label{fig1}
\end{figure}

Regression models are typically constructed to model the mean of a distribution. However, the density of the BP distribution is given in Equation (\ref{inv:01}), where it is indexed by $\alpha$ and $\beta$. In this context, in this section, we considered a new parameterization of the BP distribution in terms of the mean and precision parameters. Consider the parameterization $\mu = \alpha/(\beta-1)$ and $\phi = \beta-2$, i.e., $\alpha = \mu( 1 + \phi)$ and
$\beta = 2 + \phi$. Under this new parameterization, it follows from (\ref{inv:05}) that
$$
\textrm{E}[Y] = \mu
\quad
\textrm{and}
\quad
\textrm{Var}[Y] = \frac{\mu(1+\mu)}{\phi}.
$$
From now on, we use the notation $Y \sim \textrm{BP}(\mu, \phi)$ to indicate that $Y$ is a random variable following a BP distribution with mean $\mu$ and
precision parameter $\phi$. Note that $\textrm{V}(\mu)=\mu(1+\mu)$ is similar to the variance function of
the gamma distribution, for which the the variance has a quadratic relation with its mean.
In fact, \cite{Morris:1982aa} wrote: ``There are six basic natural exponential families having quadratic variance functions: normal, Poisson, gamma, binomial, negative binomial (geometric), and generalized hyperbolic secant distributions". However, the BP distribution can be written in the exponential family form (see Remark \ref{re01}) and has quadratic variance function (using the proposed parameterization). %Furthermore, note that $\textrm{V}(\mu)=\mu(1+\mu)$ is the variance function of geometric distribution, i.e., the variance function of BP distribution is the same of geometric distribution.
We note that this parameterization was not proposed in the statistical literature.
Using the proposed parameterization, the \text{BP} density in (\ref{inv:01}) can be written as
\begin{equation}\label{inv:011}
f(y| \mu, \phi) = \frac{y^{\mu(\phi + 1)-1}(1 + y)^{-[\mu(\phi + 1) + \phi + 2]}}{B(\mu(1+ \phi), \phi + 2)}, \quad y > 0,
\end{equation}
where $\mu > 0$ and $\phi > 0$. Figure \ref{fig1} displays some plots of the density function in (\ref{inv:011}) for some parameter values.
It is evident that the distribution is very flexible and it can be an interesting alternative to other distributions
with positive support.

The gamma (GA), reparameterized Birnbaum-Saunders (RBS) \citep{Santos-Neto:2016aa} and BP distributions have common properties such as unimodality  in the density function and quadratic variance function. %In the following, we present some motivations for the practical use of the BP distribution against the gamma and inverse Gaussian distributions.
The most important indices of the shape of a distribution are the skewness and kurtosis.
Table \ref{quantities} gives a summary of the two indices for the GA, RBS and
BP distributions. %From Table \ref{quantities}, it is possible to show that the skewness (for $\phi > 1$) and kurtosis (for $\phi > 2$) of the BP %distribution can be much larger than those of the GA and IG distributions.
\begin{table}[!t]
\centering
\caption{Skewness and kurtosis of the GA, RBS and BP distributions.}\label{quantities}
\renewcommand{\arraystretch}{1.3}
%\resizebox{\linewidth}{!}{
\begin{tabular}{lc|c}
\hline
&Skewness&Kurtosis\\ \hline
GA&$\frac{2}{\sqrt{\phi}}$&$\frac{6}{\phi} + 3$\\
%IG&$3\left(\frac{\mu}{\phi}\right)^{1/2}$&$\frac{15\,\mu}{\phi} + 3$\\
RBS &$\frac{4(3\,\phi + 11)}{(2\,\phi + 5)^{3/2}}$ &$\frac{3(41\,\phi + 186)}{(2\,\phi + 5)^{2}} + 3$\\
BP &$\frac{2(1+\phi)(1+2\,\mu)}{\phi-1}\sqrt{\frac{\phi}{\mu(1+\mu)(1+\phi)^2}}, \quad \phi > 1$ &$6\left[\frac{5\,\phi - 1}{(\phi-2)(\phi-1)} + \frac{\phi}{\mu(1+\mu)(\phi-2)(\phi-1)}\right] + 3, \quad \phi > 2$  \\
\hline
\end{tabular}
%}
\end{table}

\newpage

\section{BP regression model}\label{sec:4}

%Let $\bm{y} = (y_1, \ldots, y_n)^T$ be a random sample, where $y_i \sim \textrm{BP}(\mu_i, \phi_i)$,
%$i = 1, \ldots, n$. Suppose the mean and precision parameters of $y_i$ satisfy the following functional relations

Let $Y_1, \ldots, Y_n$ be $n$ independent random variables, where each $Y_i$, $i = 1, \ldots, n$, follows the pdf given in~\eqref{inv:011} with mean $\mu_i$ and
precision parameter $\phi_i$. Suppose the mean and the precision parameter of $Y_i$ satisfies the following functional relations
\begin{equation}\label{cs1}
g_1(\mu_i) = \eta_{1i} = \mathbf{x}^\top_i\bm{\beta} \quad \textrm{and} \quad g_2(\phi_i) = \eta_{2i} = \mathbf{z}^\top_i\bm{\nu},
\end{equation}
where $\bm{\beta} = (\beta_1, \ldots, \beta_p)^\top$ and $\bm{\nu} = (\nu_1, \ldots, \nu_q)^\top$
are vectors of unknown regression coefficients which are assumed to be functionally independent,
$\bm{\beta} \in \mathbb{R}^p$ and $\bm{\nu} \in \mathbb{R}^q$, with $p + q < n$,
$\eta_{1i}$ and $\eta_{2i}$ are the linear predictors, and $\mathbf{x}_i = (x_{i1}, \ldots, x_{ip})^\top$
 and $\mathbf{z}_i = (z_{i1}, \ldots, z_{iq})^\top$ are observations on $p$ and $q$ known regressors, for $i = 1, \ldots, n$. Furthermore, we assume that the covariate matrices $\mathbf{X} = (\mathbf{x}_1, \ldots, \mathbf{x}_n)^\top$ and $\mathbf{Z} = (\mathbf{z}_1, \ldots, \mathbf{z}_n)^\top$ have rank $p$ and $q$, respectively. The link functions $g_1: \mathbb{R} \rightarrow \mathbb{R}^+$ and $g_2: \mathbb{R} \rightarrow \mathbb{R}^+$ in (\ref{cs1}) must be strictly monotone, positive and at least twice differentiable,
such that $\mu_i = g_1^{-1}(\mathbf{x}_i^\top\,\bm{\beta})$ and $\phi_i = g_2^{-1}(\mathbf{z}_i^\top\,\bm{\nu})$, with $g_1^{-1}(\cdot)$ and
$g_2^{-1}(\cdot)$ being the inverse functions of $g_1(\cdot)$ and $g_2(\cdot)$, respectively.

%Let $\vec{y} = (y_1, \ldots, y_n)^\top$ be a random sample, where $y_i \sim \text{BP}(\mu_i, \phi)$, $i = 1, \ldots, n$.
%The beta prime regression model is defined as
%$$g(\mu_i) = \sum_{j=1}^{k}x_{ij}\beta_j = \vec{x}_i^\top\vec{\beta} = \eta_i \quad \mbox{and} \quad \textrm{Var}[Y_i] = \textrm{d}(\phi)\cdot \textrm{V}(\mu)$$
%where  $\vec{\beta} = (\beta_1, \ldots, \beta_k)^\top$ is a $k \times 1$ vector of unknown regression parameters $(k < n)$, $x_i =
%(x_{i1}, \ldots, x_{ik})^\top$ is the vector of $k$ regressors (or independent variables or covariates), such that $\mu_i = g^{-1}(x_i^\top\,\beta)$,
%with $g^{-1}(\cdot)$ being the inverse function of $g(\cdot)$, and $\eta_i$ is a linear predictor. Here, the link function
%$g:\mathbb{R}\rightarrow \mathbb{R}^{+}$ is strictly monotone, positive and at least twice differentiable.

%We define $\ell(y |\bm{\theta}) = \log[f(y| \vec{\theta}\, )]$, where $\bm{\theta}=(\mu, \phi)^\top \in (0, \infty)\times(0, \infty)$
The log-likelihood function has the form
\begin{eqnarray}\label{logm}
\ell(\bm{\beta}, \bm{\nu}) = \sum_{i=1}^{n}\ell(\mu_i, \phi_i),
\end{eqnarray}
where
\begin{eqnarray*}
\ell(\mu_i, \phi_i)&=& [\mu_i(1+\phi_i) - 1]\log(y_i) - [\mu_i(1+ \phi_i) + \phi_i +2]\log(1+y_i) -\log[ \Gamma(\mu_i(1+\phi_i))] \nonumber \\
&&- \log[ \Gamma(\phi_i+2)] + \log[ \Gamma(\mu_i(1+\phi_i) + \phi_i +2)].
\end{eqnarray*}

% \mu_i\left[\log(y_i) - \Psi^{(0)}(\mu_i(1+\phi_i))\right] -  (1+\mu_i)\left[\log(1+y_i) - \Psi^{(0)}(\mu_i(1+\phi_i) + \phi_i +2)\right] \\&-& \Psi^{(0)}(\phi_i+2) = \textrm{v}_i,

The score vector, obtained by differentiating the log-likelihood function, given in \eqref{logm}, with respect to the $\bm{\beta}$ and $\bm{\nu}$ has components
\begin{eqnarray*}
\mathbf{U}_{\bm{\beta}} = \mathbf{X}^\top \, \bm{\Phi} \, \mathbf{D}_1\, (\mathbf{y}^{*}-\bm{\mu^*})  \quad \text{and} \quad \mathbf{U}_{\bm{\nu}} = \mathbf{Z}^\top\,\mathbf{D}_2\,(\mathbf{y}^{\star}-\bm{\mu^\star}),
\end{eqnarray*}
where $\mathbf{D}_1$, $\mathbf{D}_2$, $\mathbf{y}^{*}$, $\mathbf{y}^{\star}$, $\bm{\mu}^{*}$ and
$\bm{\mu}^\star$ are given in Appendix~\ref{appA}.
% \mathbf{D}_1 = \left[a_i\, \delta_{ij}\right]$, $\mathbf{D}_2 = \left[b_i\, \delta_{ij}\right]$, $%\bm{\Phi} = \left[(1+\phi_i)\,\delta_{ij}\right]$,
%$\mathbf{y}^{*} = (y_1^{*}, \ldots, y_n^{*})\top$, $\mathbf{y}^{\star} = (y_1^{\star}, \ldots, y_n^{\star})\top$, $\bm{\mu}^{*} = (\mu_1^{*}, \ldots, \mu_n^{*})^\top$, $\bm{\mu}^\star = (\mu^\star_1, \ldots, \mu^\star_n)^\top$ and $\delta_{ij}$ is the Kronecker delta for $i,j=1, 2, \ldots, n$.

The maximum likelihood (ML) estimators $\bm{\widehat{\beta}}$ and $\bm{\widehat{\nu}}$ of $\bm{\beta}$ and $\bm{\nu}$, respectively, can be
obtained by solving simultaneously the nonlinear system of equations
$\mathbf{U}_{\bm{\beta}} = \textbf{0}$ and $\mathbf{U}_{\bm{\nu}} = \textbf{0}$. However, no closed-form expressions for the ML estimates are possible. Therefore, we must use an iterative method for nonlinear optimization. We can estimate the parameters of the
BP regression model defined in~\eqref{cs1} using the \textbf{gamlss} function, contained in \textbf{R} ~\citep[][]{r:2017} package of the same name~\citep[][]{JSSv023i07}. Theoretical results of this paper have been implemented in this  programming language. In special, the \textbf{glmBP} package provides functions for fitting BP regression models using the \textbf{gamlss} function. Log-likelihood maximizations were carried out using the RS nonlinear optimization algorithm~\citep[][]{JSSv023i07}. The current version can be downloaded from GitHub via

\begin{lstlisting}[language=R]
devtools::install_github("santosneto/glmBP")
\end{lstlisting}

This package contains a collection of routines for analyzing data from BP distribution. For example, the functions \textbf{dBP}, \textbf{pBP}, \textbf{qBP} and \textbf{rBP} implement the density function, distribution function, quantile function and random number generation for the BP distribution. Other functions are: \textbf{residuals.pearson} (Pearson residuals), {\bf envelope.BP} (simulated envelope) and \textbf{diag.BP} (local influence).

%, for example, Fisher %scoring algorithm.
%

%The next step is to obtain an expression for Fisher's information matrix.
%Consider the complete parameter vector $\bm{\theta} = (\bm{\beta}^\top, \bm{\nu}^\top)^\top$ and $\bm{\widehat\theta} = (\bm{\widehat\beta}^\top, \bm{\nu}^\top)^\top$ the ML estimators of $\bm{\theta}$.
%Let $\bm{\mathbf{F}} = \bm{\mathbf{F}}(\bm{\beta}, \bm{\nu})$ be the expected Fisher information matrix for $(\bm{\beta}, \bm{\nu})$.

When $n$ is large and under some regularity conditions \citep[see][]{cookwei:83}, we have that
\[
\left[\begin{array}{c}
\bm{\widehat{\beta}}\\
{\bm{\widehat{\nu}}}
\end{array}\right]
\build{\sim}{a}{} \mbox{N}_{p + q}\left(
\left[\begin{array}{c}
\bm{{\beta}}\\
{\bm{{\nu}}}
\end{array}\right]
, \mathbf{I}^{-1}\right),
\]
where $\build{\sim}{a}{}$ means ``approximately distributed''. Also, $\bm{\mathbf{I}}^{-1}$ may be approximated by
$(-\mathbf{H})^{-1}$, where $\mathbf{H} = \partial^2 \ell(\bm{\theta})/\partial \bm{\theta} \partial \bm{\theta}^\top(\bm{\widehat{\theta}})$, $\bm{\theta} = (\bm{\beta}^\top, \bm{\nu}^\top)^\top$, is the $(p + q) \times (p + q)$ Hessian matrix evaluated at $\bm{\widehat\theta} = (\bm{\widehat{\beta}}^\top, \bm{\widehat{\nu}}^\top)^\top$. The matrices $\mathbf{H}$ and $\mathbf{I}$ are provided in Appendix~\ref{appB}.
We then use the estimated parameters in $\mathbf{I}$ to obtain $\widehat{\mathbf{I}}$ and obtain $100(1-\gamma)\%$ asymptotic confidence intervals (CI) for $\theta_i (i=1,\ldots, p+q)$.

Consider the regression model defined in \eqref{cs1} and the corresponding log-likelihood function given in (\ref{logm}). A test of the null hypothesis that the precision is constant can be
conducted by testing the null hypothesis that $\nu_j = 0,\,\forall~j = 2, \ldots, q$ versus the alternate hypothesis that $\nu_j \neq 0$ for at least one $j~(j = 2, \ldots, q)$. Under $\mathcal{H}_0$~(null hypothesis), $n$ large and regularity conditions,
%the gradient, score, Wald and
the likelihood ratio (LR) statistic follows a $\chi^2$ distribution with $q-1$ degrees of freedom.
For testing  $\mathcal{H}_0$, we can write the LR statistic as
\begin{eqnarray}\label{lr}
\textrm{LR} &=& 2\,\sum_{i=1}^{n}\left\{ \left[ \widehat{\mu}_i(1+\widehat{\phi}_i) - \widetilde{\mu}_i(1+\widetilde{\phi}_i)\right]\log\left(\frac{y_i}{1+y_i}\right) - (\widehat{\phi} - \widetilde{\phi})\log(1+y_i)  - \log\left(\frac{\Gamma(\widehat{\mu}_i(1+\widehat{\phi}_i))}{\Gamma(\widetilde{\mu}_i(1+\widetilde{\phi}_i))} \right) \right. \nonumber \\
&&\left.- \log\left(\frac{\Gamma(\widehat{\phi}_i+2)}{\Gamma(\widetilde{\phi}_i+2)} \right) + \log\left(\frac{\Gamma(\widehat{\mu}_i(1+\widehat{\phi}_i) + \widehat{\phi}_i +2 )}{\Gamma(\widetilde{\mu}_i(1+\widetilde{\phi}_i) + \widetilde{\phi}_i +2 )} \right)\right\},
\end{eqnarray}
where $\widehat{(\cdot)}$ and $\widetilde{(\cdot)}$ denote the restricted and unrestricted maximum likelihood estimators, respectively.

\section{Diagnostic analysis}\label{sec:5}

\subsection{Residuals}

In this section, to study departures from the error assumptions as well as the presence of outlying observations, we will consider two kinds of residuals for the BP regression model defined in \eqref{cs1}.

\subsubsection{Quantile residuals}

Consider the cdf $F(y|\mu, \phi)$ of the BP distribution given in (\ref{int:01}). Since $F(\cdot)$ is continuous, then $F(y|\mu, \phi) \sim \mathcal{U}(0,1)$. Then, the quantile residual ~\citep[][]{Dunn:1996aa} under the BP regression model is defined by
\begin{equation*}
r_i^{\mathbf{Q}}=\Phi^{-1}\{F(y_i|\widehat \mu_i, \widehat\phi_i)\}, \quad i=1,2,\ldots,n,
\label{quanti}
\end{equation*}
where $\Phi(\cdot)$ is the cumulative distribution function of the standard normal distribution and $\widehat \mu$ and $\widehat \phi$ are the maximum likelihood estimators (MLE's) of $\mu$ and $\phi$. Except for the variability due to the estimators of the parameters, the distribution of the quantile residuals is standard normal.

\subsubsection{Pearson residuals}

The second residual is based on the difference $y_i - \widehat \mu_i$ and given by

\begin{equation*}
r_i^\mathbf{P} = \frac{\widehat\phi^{1/2}_i(y_i - \widehat \mu_i)}{\sqrt{ \widehat \mu_i (1+\widehat \mu_i) }},
\end{equation*}
where $\widehat \mu$ and $\widehat \phi$ are the MLE's of $\mu$ and $\phi$. For more details, see~\cite{Leiva:2014aa} and \cite{Santos-Neto:2016aa}. According to~\cite{Nelder:1972aa} a disadvantage of the Pearson residual is that the distribution for non-normal distributions
is often markedly skewed, and so it may fail to have properties similar to those of the normal-theory residual. Despite this, we will study this residual because there is no information of its performance when considering the BP regression models.

%{\color{red}
%
\subsection{Local influence}
The local influence approach based on normal curvature is recommended when the concern is related to investigate the model sensitivity under some minor perturbations in the model \citep{Cook:1986aa}. The likelihood displacement is defined as  $\textrm{LD}(\bm{\omega})=2\left[\ell(\widehat{\bm{\theta}}\,) - \ell(\widehat{\bm{\theta}}_{\omega})\right]$,  where $\widehat{\bm{\theta}}_{\omega}$ is the MLE of $\bm{\theta} = (\bm{\beta}^\top, \bm{\nu}^\top)^\top$ for a perturbed model and $\bm{\omega}= (\omega_1, \ldots,\omega_n)^{\top}$ is a perturbation vector.  \cite{Cook:1986aa} proposed to study the local behavior of $\textrm{LD}(\bm{\omega})$ around $\bm{\omega}_{0}$, which represents the null perturbation vector, such that $\textrm{LD}(\bm{\omega}_0) = 0$.

The normal curvature for $\widehat{\bm{\theta}}$ at the arbitrary direction $\bm{l}$, with $\left\|\bm{l}\right\|=1$, is given by $C_{\bm{l}}(\widehat{\bm{\theta}}\,)= 2|{\bm{l}}^{\top}{\mathbf{\Delta}}^{\top}\widehat{\mathbf{H}}^{-1}{\mathbf{\Delta}}\bm{l}|$, where $\widehat{\mathbf{H}}$ is the Hessian matrix of evaluated at $\widehat{\bm{\theta}}$ and $\mathbf{\Delta}$ is a $(p+q)\times n$ perturbation matrix with elements
\[
{\bm \Delta_{ij}}=  \pdv[2]{\ell(\bm{\theta},\bm{\omega})}{\theta_i}{\omega_j} \Big|_{\bm{
\theta} = \widehat{\bm{\theta}},\,\bm{\omega}=\bm{\omega}_0} \quad
i=1, \ldots, p + q, \, j = 1, \ldots, n,
\]
and $\ell(\bm{\theta},\bm{\omega})$ being the log-likelihood function
corresponding to the model perturbed by $\bm{\omega}$. The maximization of $C_{\bm{l}}(\widehat{\bm{\theta}}\,)$ is equivalent to finding the largest eigenvalue $C_{\textrm{max}}$
of the matrix $\widehat{\mathbf{B}} = {\mathbf{\Delta}}^{\top}\widehat{\mathbf{H}}^{-1}{\mathbf{\Delta}}$ and the direction of largest perturbation around $\bm{\omega}_{0}$, denoted by $\bm{l}_{\textrm{max}}$, is the
corresponding eigenvector. The index plot of $\bm{l}_{\textrm{max}}$ may be helpful in assessing
the influence of small perturbation on likelihood displacement.

In this work, the main idea is to study the normal curvatures for $\Hat{\bm
\beta}$ and $\Hat{\bm
\nu}$ in the unitary direction $\bm{l}$ available at  $\widehat{\bm{\theta}}$ and $\omega_0$. Such curvatures are expressed as
$C_{\bm l}(\Hat{\bm \beta})= 2|{\bm l}^{\top}{\bm
\Delta}^{\top}[\widehat{\mathbf{H}}^{-1} - \Hat{\mathbf{H}}_\nu]{\bm \Delta}{\bm l}|$, where
\begin{equation*}
\Hat{\mathbf{H}}_\nu
=\left[
\begin{array}{cc}
{\bm 0}  & {\bm 0}  \\
{\bm 0}  & \Hat{\mathbf{H}}_{\nu\nu}^{-1} \\
\end{array}
\right] \label{inflm}
\end{equation*}
in case the interest is only  on the vector $\Hat{\bm
\beta}$, and
$C_{\bm l}(\Hat{\bm \nu})= 2|{\bm l}^{\top}{\bm
\Delta}^{\top}[\widehat{\mathbf{H}}^{-1} - \Hat{\mathbf{H}}_\beta]{\bm \Delta}{\bm l}|$, where
\begin{equation*}
\Hat{\mathbf{H}}_\beta
=\left[
\begin{array}{cc}
\Hat{\mathbf{H}}_{\beta\beta}^{-1} & {\bm 0}  \\
{\bm 0}  &{\bm 0}  \\
\end{array}
\right], \label{inflm}
\end{equation*}
if the interest lies in studying the local influence on $\Hat{\bm \nu}$.
To assess the influence of the perturbations we may consider the index plot for the normal curvature $C_i(\widehat{{\bm \theta}}) = 2|\hat{b}_{ii}|$, where $\hat{b}_{ii}$ is the $i$th diagonal element of $\widehat{\mathbf{B}}$. \citet{lv:98} suggested to pay a special attention in those observations with $C_i > 2\,\overline{C}$, where $\overline{C}= (1/n)\sum_{i=1}^{n} C_i$.

In this paper, we calculate the matrix $\mathbf{\Delta}$ for five different perturbation schemes, namely: case weighting perturbation, response perturbation, mean covariate perturbation, precision covariate perturbation, and simultaneous covariate perturbation. These results are presented with
details in Appendix~\ref{appC}. Figure~\ref{diagram} presents a diagram with the perturbation matrices under different schemes.
\tikzstyle{decision} = [diamond, draw, fill=white,
    text width=4.5em, text badly centered, node distance=3cm, inner sep=0pt]
\tikzstyle{block} = [rectangle, draw, fill=white,
    text width=5em, text centered, rounded corners, minimum height=5em]
\tikzstyle{line} = [draw, -latex']
\tikzstyle{cloud} = [rectangle, draw, fill=white,
    text width=10em, text centered, rounded corners, minimum height=5em]
\tikzstyle{cloudb} = [rectangle, draw, fill=white,
    text width=20em, text centered, rounded corners, minimum height=20em]

\begin{figure}[t]
\centering
\begin{tikzpicture}[node distance = 6.cm, auto]
    %Place nodes
    \node [block] (init) {Perturbation schemes};
    \node [cloud, left of=init] (expert){
$
\widehat{\mathbf{\Delta}} = \left[\begin{array}{c}
\mathbf{X}^{\top}\, \widehat{\mathbf{D}}_1\, \widehat{\mathbf{D}}_6   \\
\mathbf{Z}^{\top}\, \widehat{\mathbf{D}}_2\, \widehat{\mathbf{D}}_7
\end{array}
\right]
$};
    \node [cloudb, below of=init] (system) {
\textbf{Mean covariate perturbation}
    \[
\widehat{\mathbf{\Delta}} =
\begin{bmatrix}
s_i\,\widehat\beta_{t}\,\mathbf{X}^\top\, \widehat{\mathbf{D}}_3 + s_i\, \widehat{\mathbf{T}}_\mu\\
s_i\,\widehat\beta_{t}\,\mathbf{Z}^\top \, \widehat{\mathbf{D}}_5 \\
\end{bmatrix}
\]

\textbf{Precision covariate perturbation}
\[
\widehat{\mathbf{\Delta}} =
\begin{bmatrix}
\dot{s}_i\,\widehat\nu_{k}\,\mathbf{Z}^\top\, \widehat{\mathbf{D}}_4 + \dot{s}_i\, \widehat{\mathbf{T}}_\phi\\
\dot{s}_i\,\widehat\nu_{k}\,\mathbf{X}^\top \, \widehat{\mathbf{D}}_5 \\
\end{bmatrix}
\]

\textbf{Simultaneous (mean and precision) covariate perturbation}
\[
\widehat{\mathbf{\Delta}} =
\begin{bmatrix}
s_i\,\left\{\mathbf{X}^\top\,\left[\widehat\beta_{t}\, \widehat{\mathbf{D}}_3 + \widehat\nu_k\,\widehat{\mathbf{D}}_5 \right] + \widehat{\mathbf{T}}_\mu\right\} \\
\\
s_i\,\left\{\mathbf{Z}^\top\,\left[\widehat\nu_{k}\, \widehat{\mathbf{D}}_4 + \widehat\beta_t\,\widehat{\mathbf{D}}_5 \right] + \widehat{\mathbf{T}}_\phi\right\} \\
\end{bmatrix}
\]

};
    \node [cloud, right of=init] (identify) {$
\widehat{\mathbf{\Delta}} = \left[\begin{array}{c}
\mathbf{X}^{\top}\, \widehat{\mathbf{D}}_1\, \widehat{\mathbf{D}}_8 \, \mathbf{S}  \\
\mathbf{Z}^{\top}\, \widehat{\mathbf{D}}_2\, \widehat{\mathbf{D}}_9 \, \mathbf{S}
\end{array}
\right]
$};
   % \node [block, below of=identify] (evaluate) {evaluate candidate models};
   % \node [block, left of=evaluate, node distance=3cm] (update) {update model};
   % \node [decision, below of=evaluate] (decide) {is best candidate better?};
   % \node [block, below of=decide, node distance=3cm] (stop) {stop};
    %Draw edges
    \path [line] (init) -- node {Response}(identify);
    \path [line] (init) -- node {Case-weights}(expert);
    \path [line] (init) -- node {Regressor}(system);
    %\path [line] (identify) -- (evaluate);
    %\path [line] (evaluate) -- (decide);
    %\path [line] (decide) -| node [near start] {yes} (update);
    %\path [line] (update) |- (identify);
    %\path [line] (decide) -- node {no}(stop);
    %\path [line,dashed] (expert) -- (init);
    %\path [line,dashed] (system) -- (init);
    %\path [line,dashed] (system) |- (evaluate);
\end{tikzpicture}
\caption{Perturbation matrices under different schemes.}\label{diagram}
\end{figure}
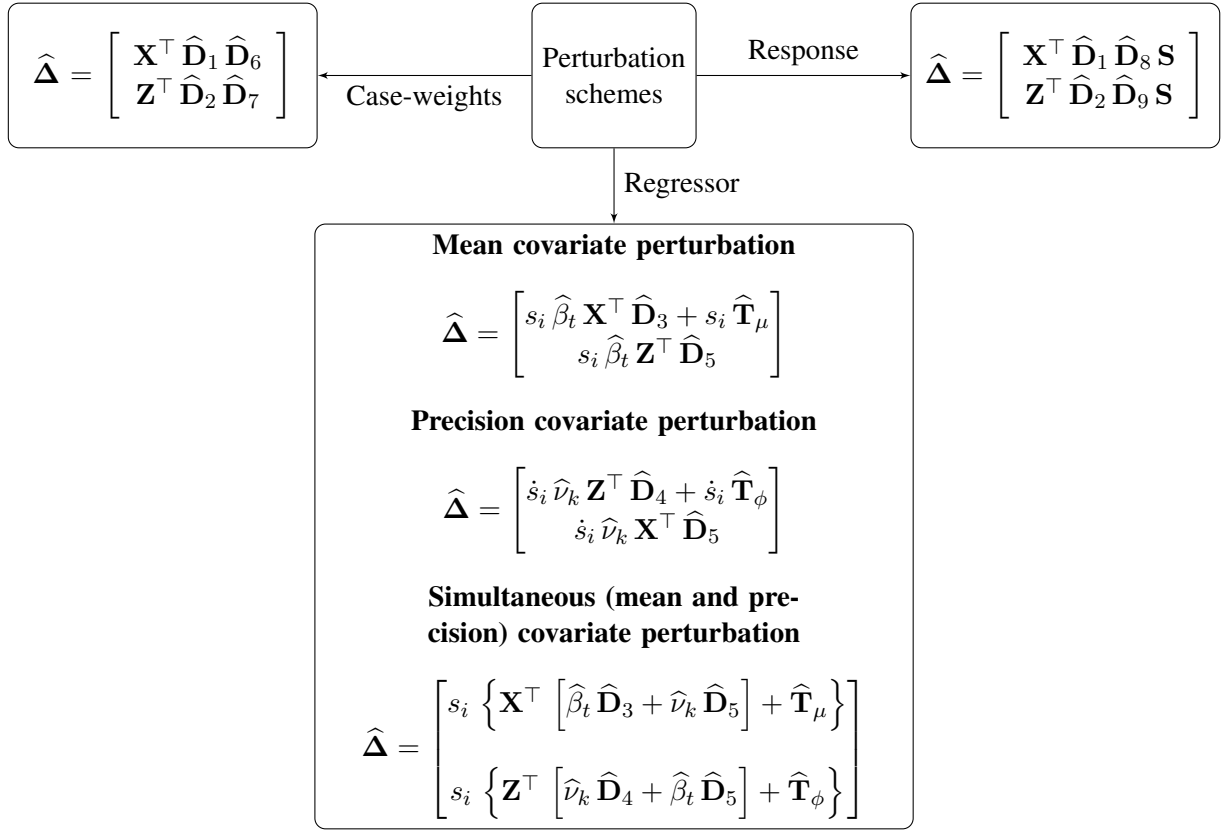
\section{Monte Carlo studies}\label{sec:6}
In this section, we present two simulation scenarios: (i) The first batch of Monte Carlo experiments is conducted to assess the finite sample behavior of the MLE's; and (ii) We perform a simulation study to examine the distributions of the residuals $r_i^{\mathbf{Q}}$ and $r_i^{\mathbf{P}}$.

For scenarios I and II, the Monte Carlo experiments were carried out using
\begin{equation}\label{modsim}
\log(\mu_i) = 2.0 - 1.6\,x_{i1} \quad \text{and}\quad \log(\phi_i) = 2.6 - 2.0\, z_{i1} \quad i = 1, \ldots, n,
\end{equation}
where the covariates $x_{i1}$ and $z_{i1}$, for $i = 1, \ldots, n$ were generated from a standard uniform.
The values of all regressors were kept constant during the simulations.
The number of Monte Carlo replications was 5,000. All simulations and all the graphics were performed
using the \textbf{R} programming language \citep[][]{r:2017}. Regressions were estimated for each sample using the \textbf{gamlss()} function.

\subsection{Scenario I}
The goal of this simulation experiment is to examine the finite sample behavior of the MLE's.
This was done 5,000 times each for sample sizes ($n$) of 50, 100 and 150. In order to analyze the point estimation results, we computed, for each sample size and for each estimator: mean (E), bias (B) and root mean square error (RMSE). We will use Monte Carlo simulations to estimate the empirical coverage probability of the asymptotic confidence intervals.	

\begin{table}[H]
\centering
\caption{Mean, bias and mean square error.} \vspace{0.1cm}
\renewcommand{\arraystretch}{1.3}
\resizebox{\linewidth}{!}{
\begin{tabular}{ccc ccc c| cccccc} \hline
\multirow{2}*{$n$}& \multicolumn{6}{c|}{\textbf{Mean parameter}} & \multicolumn{6}{c}{\textbf{Precision parameter}} \\ \cline{2-13}
 &$\textrm{E}(\widehat{\beta}_0)$ &$\textrm{E}(\widehat{\beta}_1)$  & $\textrm{B}(\widehat{\beta}_0)$  &$\textrm{B}(\widehat{\beta}_1)$  & $\textrm{RMSE}(\widehat{\beta}_0)$ & $\textrm{RMSE}(\widehat{\beta}_1)$ &$\textrm{E}(\widehat{\nu}_0)$ & $\textrm{E}(\widehat{\nu}_1)$& $\textrm{B}(\widehat{\nu}_0)$ &$\textrm{B}(\widehat{\nu}_1)$ & $\textrm{RMSE}(\widehat{\nu}_0)$ &$\textrm{RMSE}(\widehat{\nu}_1)$ \\ \hline
50&1.996 &$-$1.601& $-$0.004& $-$0.001& 0.114& 0.226 & 2.742& $-$2.117&  0.142& $-$0.117&0.776& 1.019\\
100&1.999 &$-$1.602& $-$0.001& $-$0.002& 0.077& 0.155& 2.661& $-$2.048&  0.061& $-$0.048&0.654& 0.636\\
150&1.998& $-$1.601& $-$0.002& $-$0.001&  0.063& 0.122& 2.642& $-$2.025 & 0.042& $-$0.025& 0.281& 0.508\\ \hline
\end{tabular}
}
\label{tab:1}
\end{table}

%Heter =  7.254188
%         B0     B1 biasB0 biasB1 varB0 varB1 eqmB0 eqmB1
%n=50  1.996 -1.601 -0.004 -0.001 0.116 0.225 0.013 0.051
%n=100 1.999 -1.602 -0.001 -0.002 0.080 0.155 0.006 0.024
%n=150 1.998 -1.601 -0.002 -0.001 0.064 0.124 0.004 0.015
%         N0     N1 biasN0 biasN1 varN0 varN1 eqmN0 eqmN1
%n=50  2.742 -2.117  0.142 -0.117 0.763 1.012 0.602 1.038
%n=100 2.661 -2.048  0.061 -0.048 0.651 0.635 0.428 0.405
%n=150 2.642 -2.025  0.042 -0.025 0.279 0.508 0.079 0.258
%      B0_nc1 B0_nc5 B0_nc10 B1_nc1 B1_nc5 B1_nc10
%n=50   0.982  0.927   0.872  0.982  0.935   0.875
%n=100  0.985  0.941   0.885  0.986  0.940   0.888
%n=150  0.986  0.942   0.889  0.988  0.944   0.890
%      N0_nc1 N0_nc5 N0_nc10 N1_nc1 N1_nc5 N1_nc10
%n=50   0.975  0.919   0.865  0.986  0.941   0.882
%n=100  0.984  0.937   0.882  0.989  0.942   0.887
%n=150  0.985  0.937   0.888  0.988  0.947   0.891
%      [,1]  [,2]  [,3]  [,4]
%[1,] 0.110 0.216 3.852 4.291
%[2,] 0.077 0.150 2.713 2.981
%[3,] 0.062 0.122 0.271 0.491

\begin{table}[H]
\centering
\caption{Standard errors and standard desviation estimates.} \vspace{0.1cm}
\renewcommand{\arraystretch}{1.0}
%\resizebox{\linewidth}{!}{
\begin{tabular}{ccc cc| cccc} \hline
\multirow{2}*{$n$}& \multicolumn{4}{c|}{\textbf{Mean parameter}} & \multicolumn{4}{c}{\textbf{Precision parameter}} \\ \cline{2-9}
 &$\textrm{SD}(\widehat{\beta}_0)$ &$\textrm{SD}(\widehat{\beta}_1)$  & $\textrm{EP}(\widehat{\beta}_0)$  &$\textrm{EP}(\widehat{\beta}_1)$   &$\textrm{SD}(\widehat{\nu}_0)$ & $\textrm{SD}(\widehat{\nu}_1)$& $\textrm{EP}(\widehat{\nu}_0)$ &$\textrm{EP}(\widehat{\nu}_1)$  \\ \hline
50&0.116 &0.225 & 0.110 &0.216& 0.763 & 1.012 & 3.852& 4.291\\
100&0.080 &0.155& 0.077& 0.150& 0.651& 0.635& 2.713&  2.981\\
150&0.064&0.124 & 0.062& 0.122&  0.279& 0.508&  0.271& 0.491\\ \hline
\end{tabular}
%}
\label{tab:se}
\end{table}

Table~\ref{tab:1} presents the mean, bias and RMSE for the maximum likelihood estimators of $\beta_0$, $\beta_1$, $\nu_0$ and $\nu_1$. The estimates of the regression parameters $\beta_0$ and $\beta_1$ are more accurate than those of $\nu_0$ and $\nu_1$. We note that the RMSEs tend to decrease as larger sample sizes are used, as expected. Finally, note that the standard deviations (SD) of the estimates very close to the asymptotic standard errors (SE) estimates when $n$ tends towards infinity (see, Table~\ref{tab:se}).

\begin{table}[H]
\centering
\caption{Coverage probabilities of asymptotic confidence intervals.} \vspace{0.1cm}
\renewcommand{\arraystretch}{1.3}
\resizebox{\linewidth}{!}{
    \begin{tabular}{ccc ccc  c| cccccc} \hline
    \multirow{2}*{$n$}& \multicolumn{6}{c|}{\textbf{Mean parameter}} & \multicolumn{6}{c}{\textbf{Precision parameter}} \\ \cline{2-13}
   &  $\textrm{CI}(\beta_0; 99\%)$& $\textrm{CI}(\beta_0;95\%)$ & $\textrm{CI}(\beta_0;90\%)$& $\textrm{CI}(\beta_1; 99\%)$& $\textrm{CI}(\beta_1; 95\%)$& $\textrm{CI}(\beta_1; 90\%)$ &   $\textrm{CI}(\nu_0; 99\%)$& $\textrm{CI}(\nu_0; 95\%)$ & $\textrm{CI}(\nu_0; 90\%)$& $\textrm{CI}(\nu_1; 99\%)$& $\textrm{CI}(\nu_1; 95\%)$ & $\textrm{CI}(\nu_1; 90\%)$ \\  \hline
50&   0.982&  0.927&   0.872&  0.982&  0.935&   0.875 &  0.975&  0.919&   0.865&  0.986&  0.941 &  0.882\\
100 & 0.985 & 0.941 &  0.885&  0.986&  0.940 &  0.888& 0.984&  0.937&   0.882&  0.989&  0.942 &  0.887\\
150 & 0.986 & 0.942 &  0.889&  0.988&  0.944 &  0.890&  0.985&  0.937&   0.888 & 0.988 & 0.947&   0.891\\
\hline
\end{tabular}
}
\label{tab:2}
\end{table}

Table~\ref{tab:2} presents the empirical coverage probabilities of the asymptotic confidence intervals of level $100(1 -
\gamma)\%$, denoted by $\textrm{CI}(\cdot; 100(1-\gamma)\%)$. Note that the asymptotic confidence intervals have an empirical coverage probability that is less than the nominal values (0.99, 0.95 and 0.90). Overall, we observe that the asymptotic confidence intervals have a good performance.

\subsection{Scenario II}
The second simulation study was performed  to examine how well the distributions of the $r_i^\mathbf{Q}$ and $r_i^\mathbf{P}$ are approximated by the standard normal distribution. %We use a BP regression
In each of the 5,000 replications, we obtain 150 observations from the BP distribution and we fit
the model given in~\eqref{modsim} and generate the residuals.
\begin{figure}[h!]
\centering
\subfigure[][\label{fig5a}]{\includegraphics[height=5.5cm,width=8.5cm]{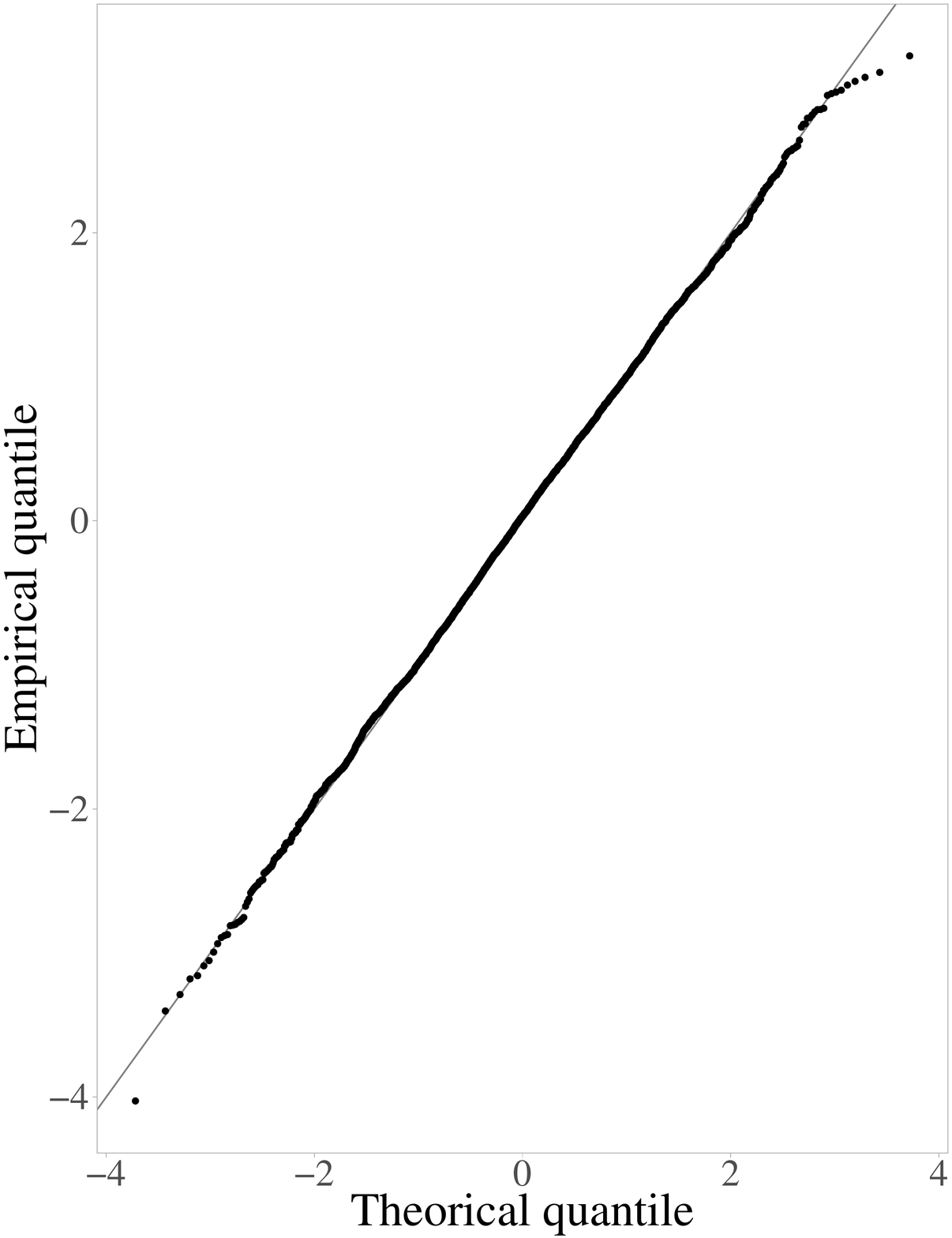}} \quad
\subfigure[][\label{fig5b}]{\includegraphics[height=5.5cm,width=8.5cm]{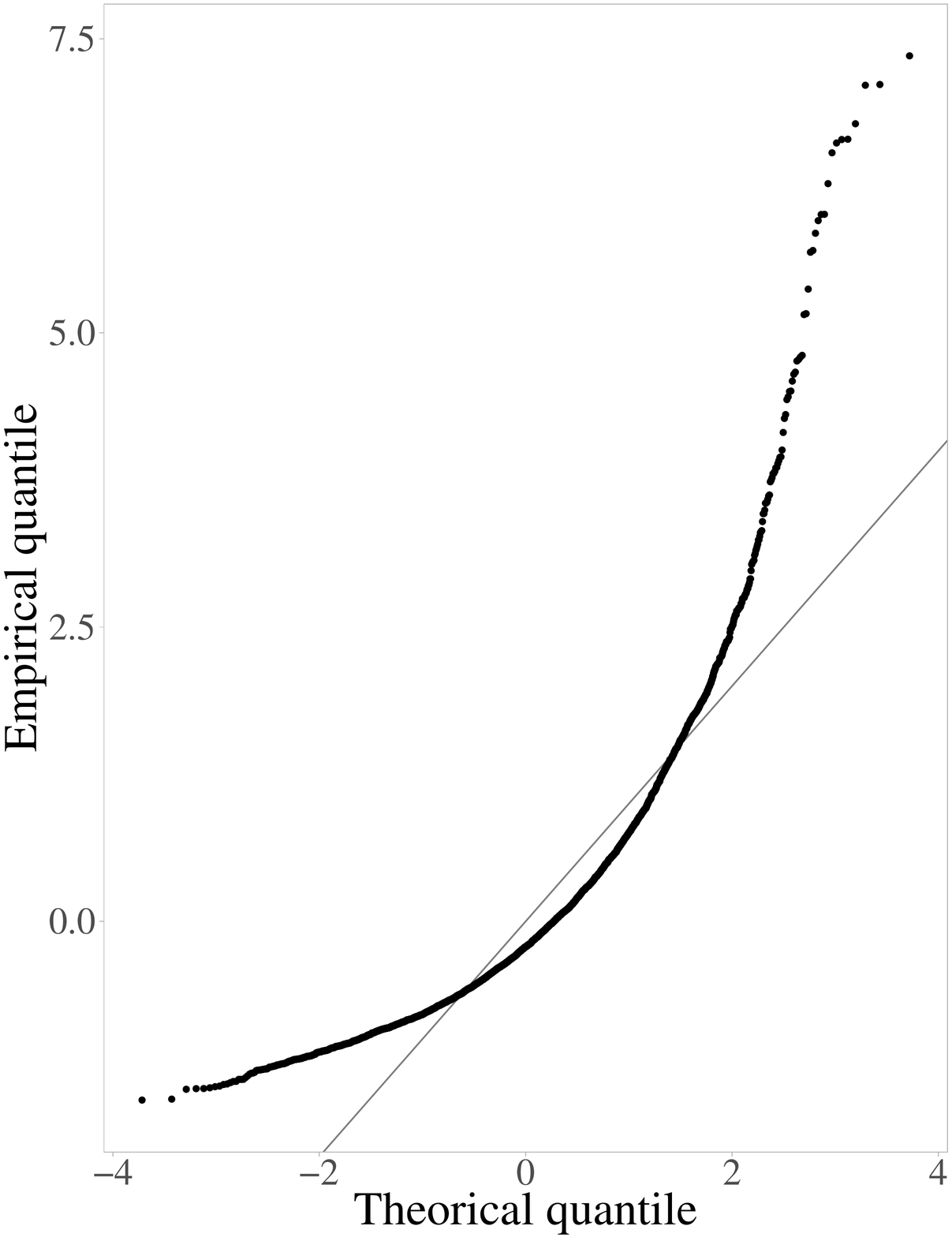}}
%\subfigure[$\phi = 100$]{\includegraphics[height=5.5cm,width=5.5cm]{density3}}
\caption{Quantile-quantile plots of the vector of residuals $r_i^\mathbf{Q}$(a) and $r_i^\mathbf{P}$(b) from the BP model.}
\label{fig5}
\end{figure}

In Figure~\ref{fig5}, we perform a comparison between the empirical distribution of the
residuals and the standard normal distribution, by using quantile-quantile plots. The residuals $r_i^\mathbf{Q}$'s have a good agreement with the a standard normal distribution. However, note that the residuals
$r_i^\mathbf{P}$'s have a distribution skewed to the right.  The results presented in Figure~\ref{fig5}  show that the distribution of the $r_i^\mathbf{Q}$'s is better approximated by the standard normal distribution than that of the $r_i^\mathbf{P}$'s .

%\subsection{Scenario IV}
%
%In this section, we develop Monte Carlo simulations to compare the performance of the precision tests presented in Section~\ref{sec:4}. We estimate the parameters of the BP regression model using the \textbf{glm.BP} package, see~Section~\ref{sec:3}. To measure the heterogeneity degree of the data, we consider the following measure
%\[
%\mathbb{H} = \frac{\max\{\phi_1,\phi_2,\ldots,\phi_n\}}{\min\{\phi_1,\phi_2,\ldots,\phi_n\}}.
%\]
%%
%Note that this is equal to 1 if and only if the BP model has precision constant. In simulations we calculate the rejection rates of the each test and the empirical power.
%
% \begin{enumerate}
% \item Empirical size:
% \item Empirical power:
% \end{enumerate}
%
%\begin{figure}[H]
%\centering
%\subfigure[][\label{fig7a}]{\includegraphics[height=6.5cm,width=8.5cm]{resQ}} \quad
%\subfigure[][\label{fig7b}]{\includegraphics[height=6.5cm,width=8.5cm]{resP}}
%%\subfigure[$\phi = 100$]{\includegraphics[height=5.5cm,width=5.5cm]{density3}}
%\caption{Empirical distributions of the residuals $r_i^\mathbf{Q}$(a) and $r_i^\mathbf{P}$(b).}
%\label{test}
%\end{figure}

\section{Real data application}\label{sec:7}
\begin{figure}[H]
\centering
\subfigure[][\label{fig6a}]{\includegraphics[height=5.5cm,width=8.5cm]{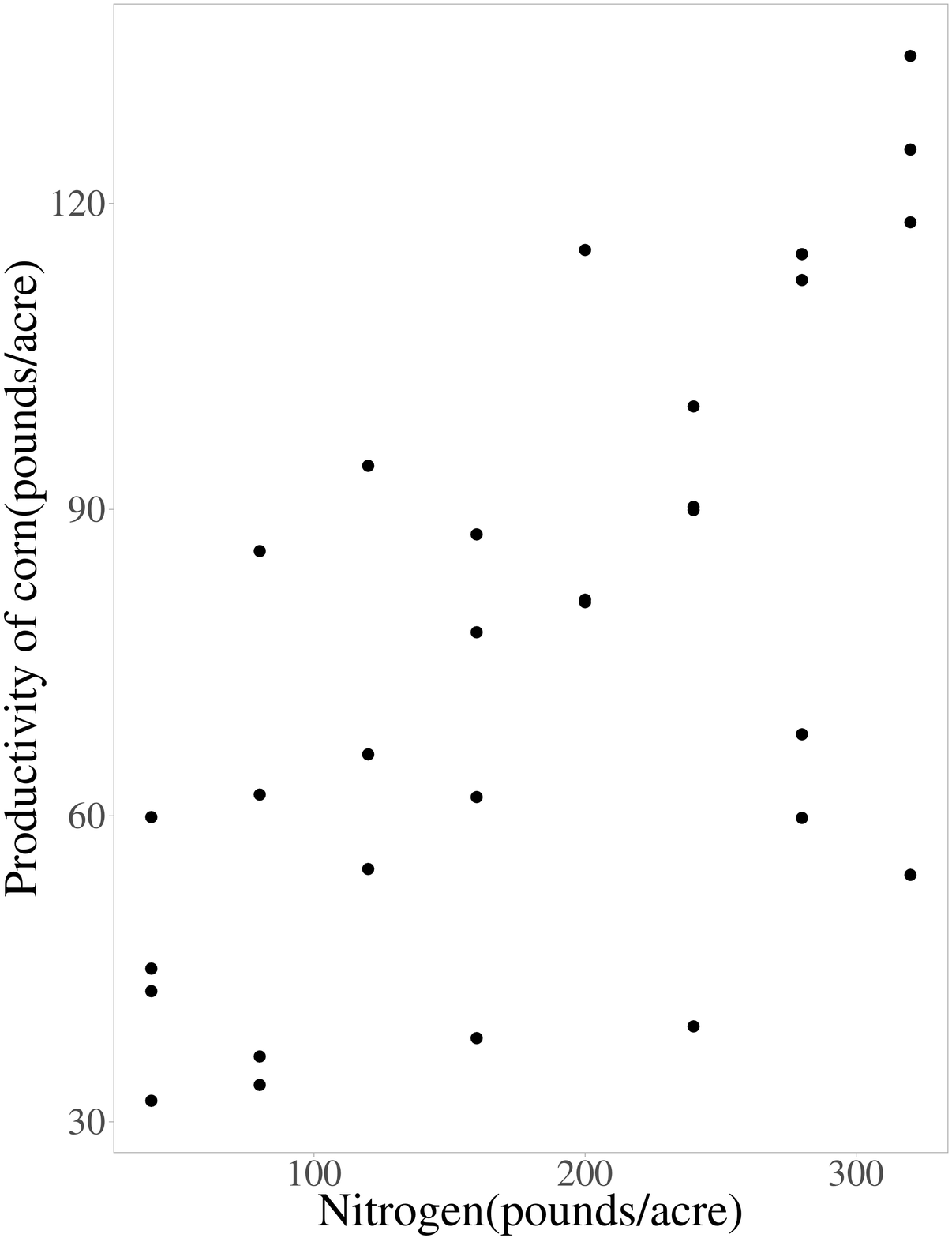}} \quad
\subfigure[][\label{fig6b}]{\includegraphics[height=5.5cm,width=8.5cm]{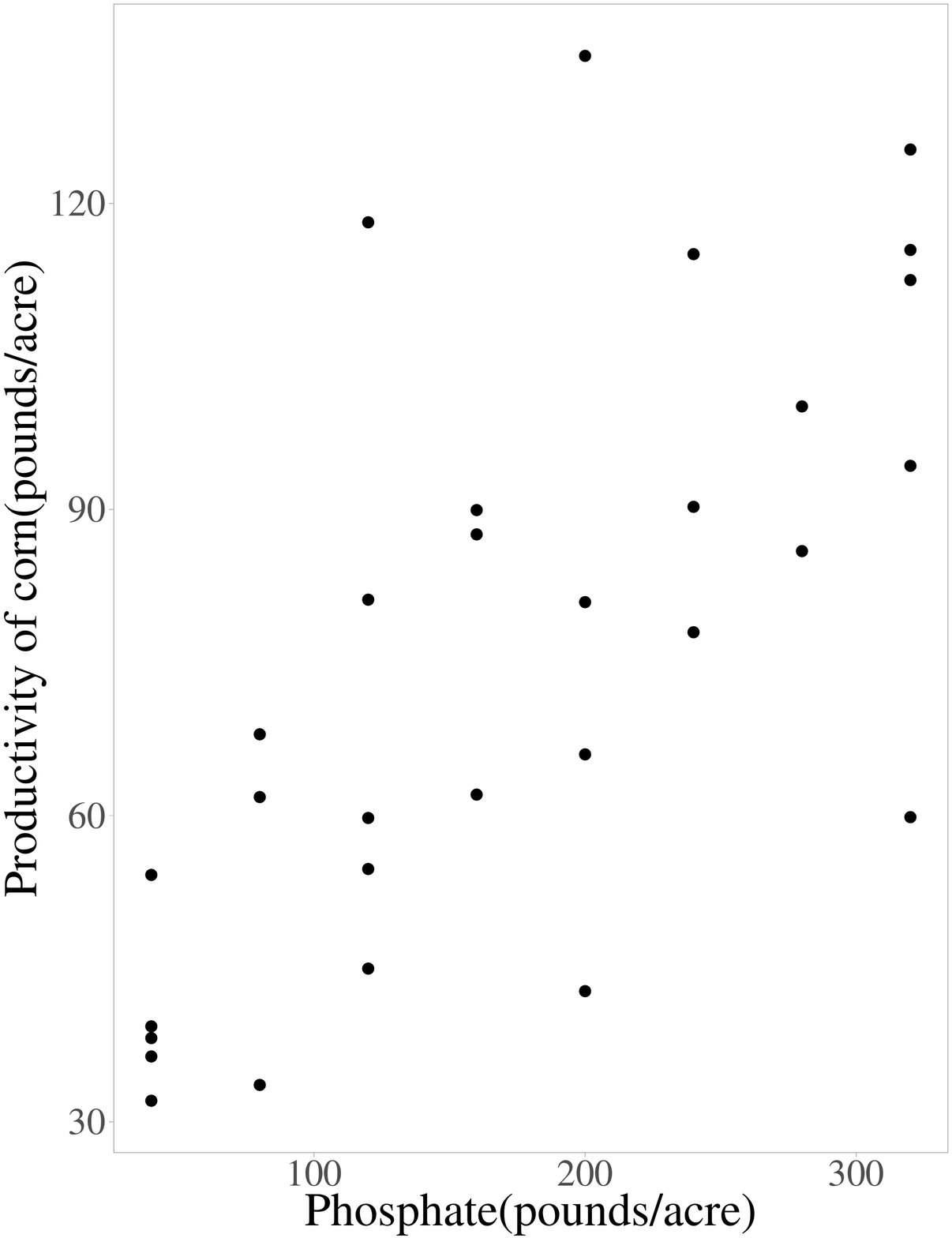}}
%\subfigure[$\phi = 100$]{\includegraphics[height=5.5cm,width=5.5cm]{density3}}
\caption{Scatterplots of nitrogen against productivity (a) and phosphate against productivity (b).}
\label{fig6}
\end{figure}

The analysis was carried out using
the \textbf{glmBP} and \textbf{gamlss} packages. We will consider a randomized experiment described in \cite{wgj:93}. In this study, the productivity of corn (pounds/acre) is studied considering different combinations of nitrogen contents and phosphate contents (40, 80, 120, 160, 200, 240, 280 and 320 pounds/acre). The response variable $Y$ is the productivity of corn given the combination of nitrogen $(x_{1i})$ and phosphate $(x_{2i})$ corresponding to the $i$th experimental condition $(i = 1, \ldots, 30)$ and the data are presented in Figure \ref{fig6}.
% These data were analyzed by Paula (2013) using the gamma model.

In Figure \ref{fig6}, there is evidence of an increasing productivity trend with increased inputs. Moreover, note that there is increased variability with increasing amounts of nitrogen and phosphate, suggesting that the assumption of GA or RBS distributions (both with quadratic variance) for $\log(\mu_i)$, i.e., we consider that $Y_i$ follows $\textrm{BP}(\mu_i, \phi_i)$, $\textrm{GA}(\mu_i, \phi_i)$ and $\textrm{RBS}(\mu_i, \phi_i)$  distributions with a systematic component given by
%Figures~\ref{fig6a}-\ref{fig6b} there is evidence of an increasing productivity trend with increased inputs. Moreover, note that there is increased variability %with increasing amounts of nitrogen and phosphate. In this way, models that can carry heteroscedasticity are good candidates. Here, we
%consider three models (gamma, inverse Gaussian and beta prime) with this characteristic. First, we consider that $Y_i \sim \mathcal{F}(\mu_i,\phi_i)$ with a %systematic component given by
%
\[
\log(\mu_i) = \beta_0 + \beta_1 \log(x_{1i}) + \beta_2 \log(x_{2i})
\quad
\textrm{and}
\quad
\phi_i = \nu_0.
\]
We compared the BP regression model with the GA and RBS regression models.

\begin{figure}[t]
\centering
\subfigure[\label{fig6a}][\textbf{BP}]{\includegraphics[height=5.5cm,width=5.5cm]{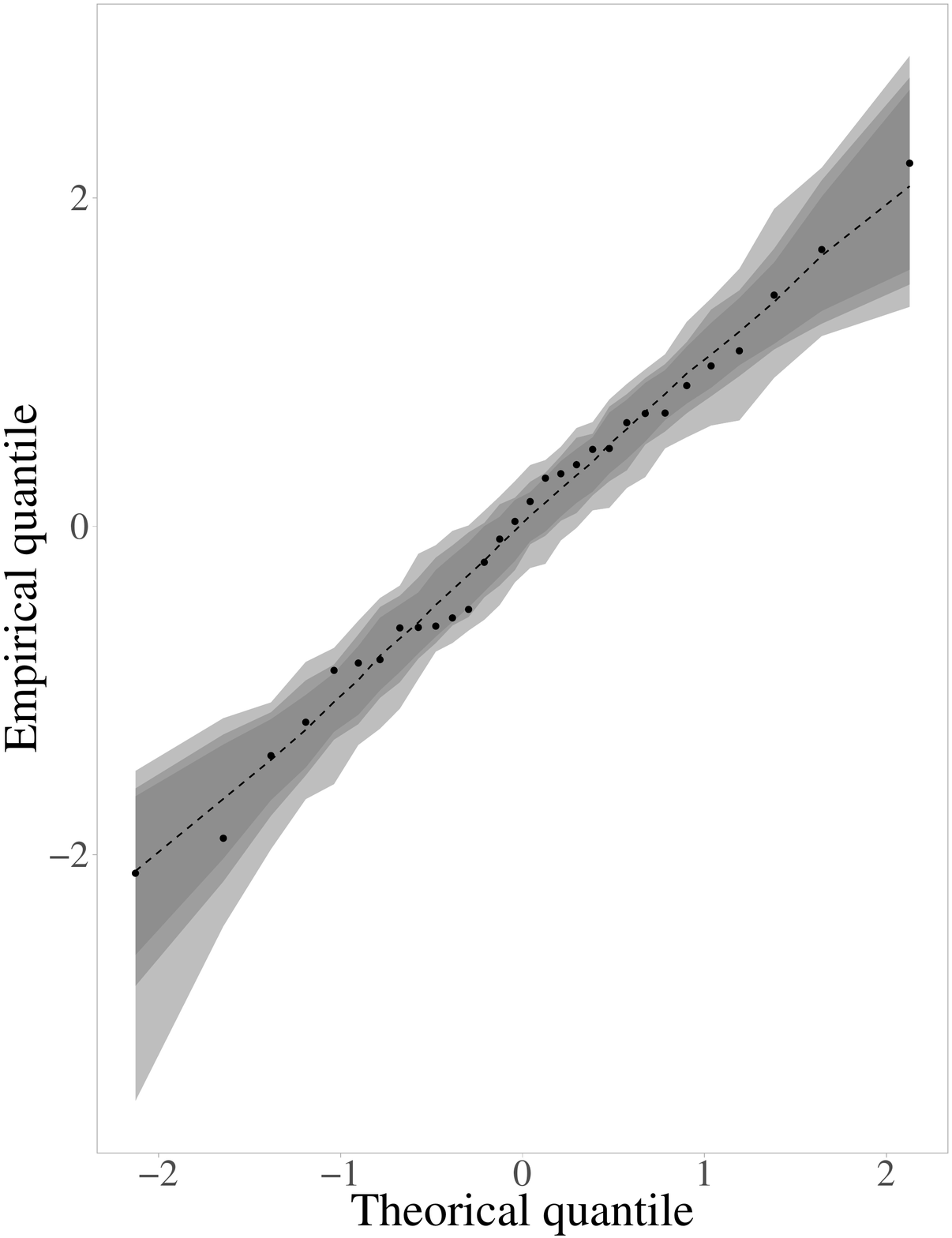}} \quad
\subfigure[\label{fig6b}][\textbf{GA}]{\includegraphics[height=5.5cm,width=5.5cm]{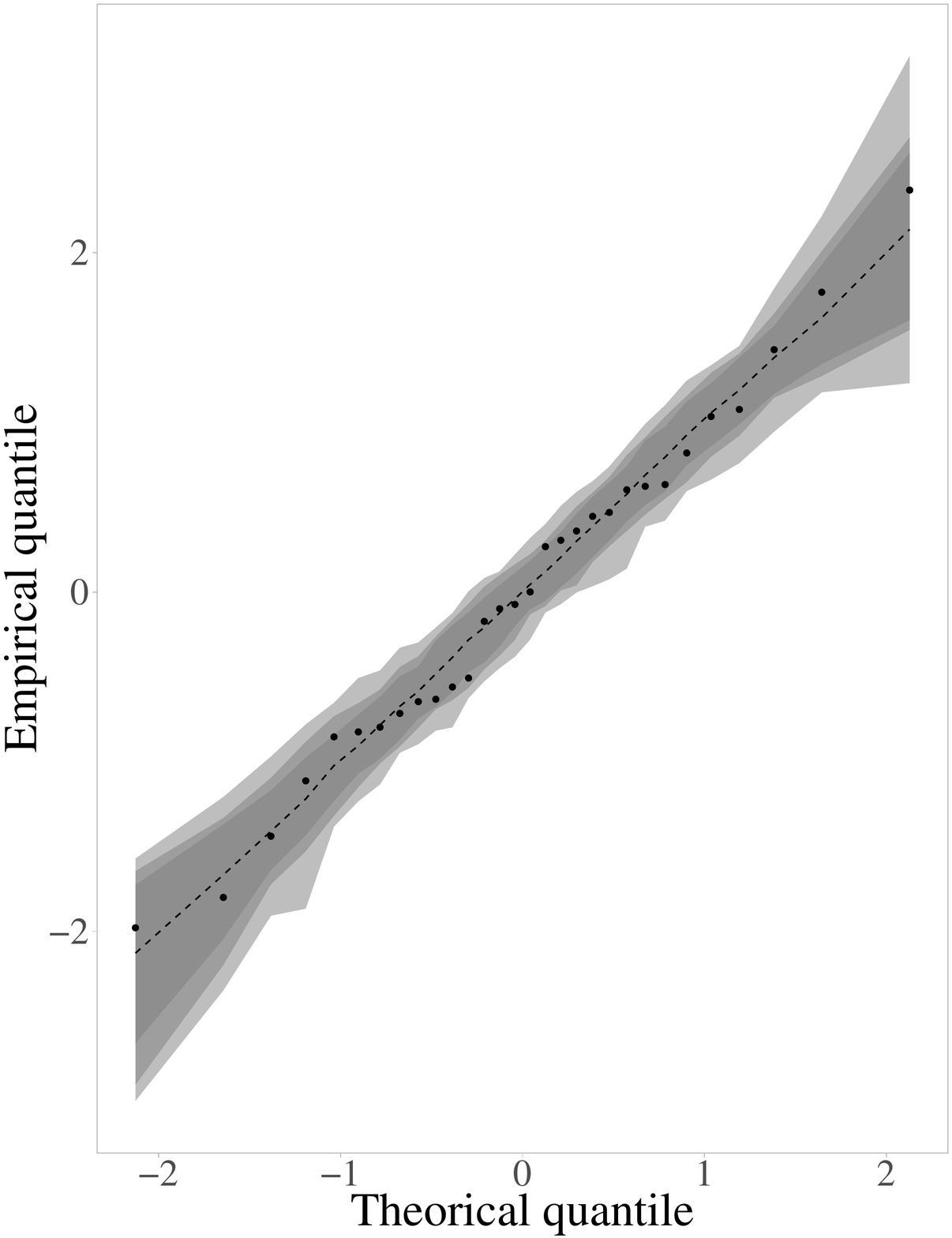}}
\subfigure[\label{fig6c}][\textbf{RBS}]{\includegraphics[height=5.5cm,width=5.5cm]{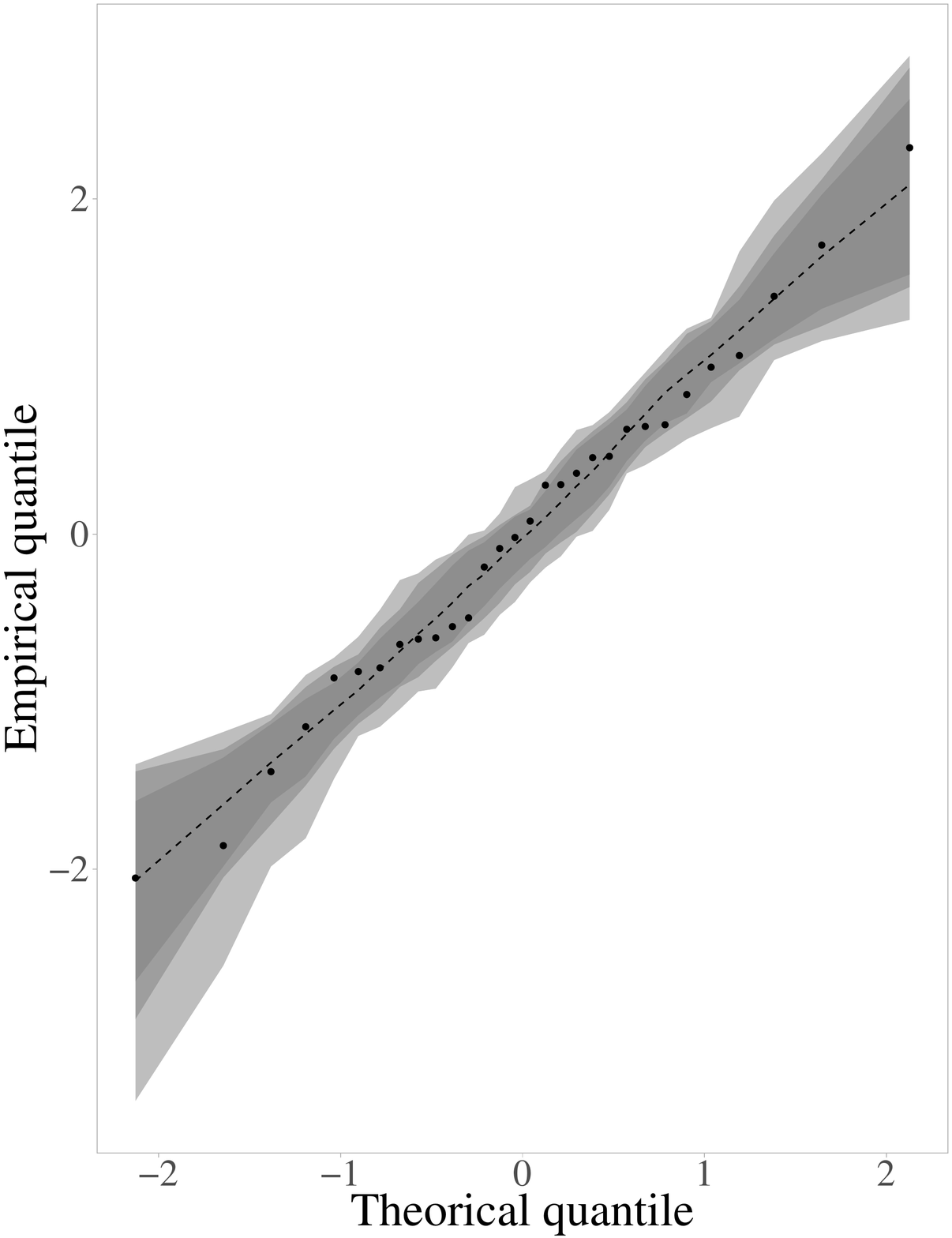}}
\caption{Simulated envelopes for the residuals $r_i^\mathbf{Q}$ from BP, GA and RBS models.}
\label{fig7}
\end{figure}

Table~\ref{aplic1} presents the estimates, standard errors (SE), Akaike information criterion (AIC) and Bayesian information criterion (BIC) for the BP, GA and RBS models. We can note that the BP, GA and RBS regression models present a similar fit according to the information criteria (AIC and BIC) used. In Figure~\ref{fig7}, we see that most of the residuals lie inside the envelopes and that there is no noticeable pattern in the residuals. Visual inspection of the plots in Figure 6 suggest that there is no serious violation of the distributional assumptions.

\begin{table}[t]
\centering
\caption{Parameter estimates and SE for the BP, GA and RBS models.}\label{aplic1}
\renewcommand{\arraystretch}{1.3}
%\resizebox{\linewidth}{!}{
\begin{tabular}{l r r r r r r r r }
\hline
\multirow{2}*{Parameter} &\multicolumn{2}{c}{\textbf{BP}}&&\multicolumn{2}{c}{\textbf{GA}}&&\multicolumn{2}{c}{\textbf{RBS}}\\ \cline{2-3} \cline{5-6} \cline{8-9}
          & Estimate& SE         &&  Estimate& SE           &&  Estimate& SE\\ \hline
$\beta_0$ &  0.4471& 0.2697   &&0.4687&0.2808 &&0.4589&0.2644 \\
$\beta_1$ &  0.3453&  0.0399   &&0.3499 &0.0421 &&0.3473 &0.0397  \\
$\beta_2$ & 0.4191 &  0.0382   &&0.4100& 0.0407&& 0.4146 &0.0384 \\
$\nu_0$    &45.650 &    12.260             &&    46.592  & 11.987 &&92.570&23.900\\ \hline
 \multicolumn{9}{c}{\textbf{Selection criteria}} \\ \hline
Log-likelihood   &   \multicolumn{2}{c}{$-$112.19}           &  & \multicolumn{2}{c}{$-$112.30}       & &\multicolumn{2}{c}{$-$113.86}\\
AIC     & \multicolumn{2}{c}{232.38}           &  & \multicolumn{2}{c}{232.59}       & &\multicolumn{2}{c}{232.41}\\
BIC    &  \multicolumn{2}{c}{237.99}           &  & \multicolumn{2}{c}{238.20}       & &\multicolumn{2}{c}{238.02}\\
\hline
\end{tabular}
%}
\end{table}

\begin{figure}[H]
\centering
\subfigure[][\textbf{BP}]{\includegraphics[height=5.5cm,width=5.5cm]{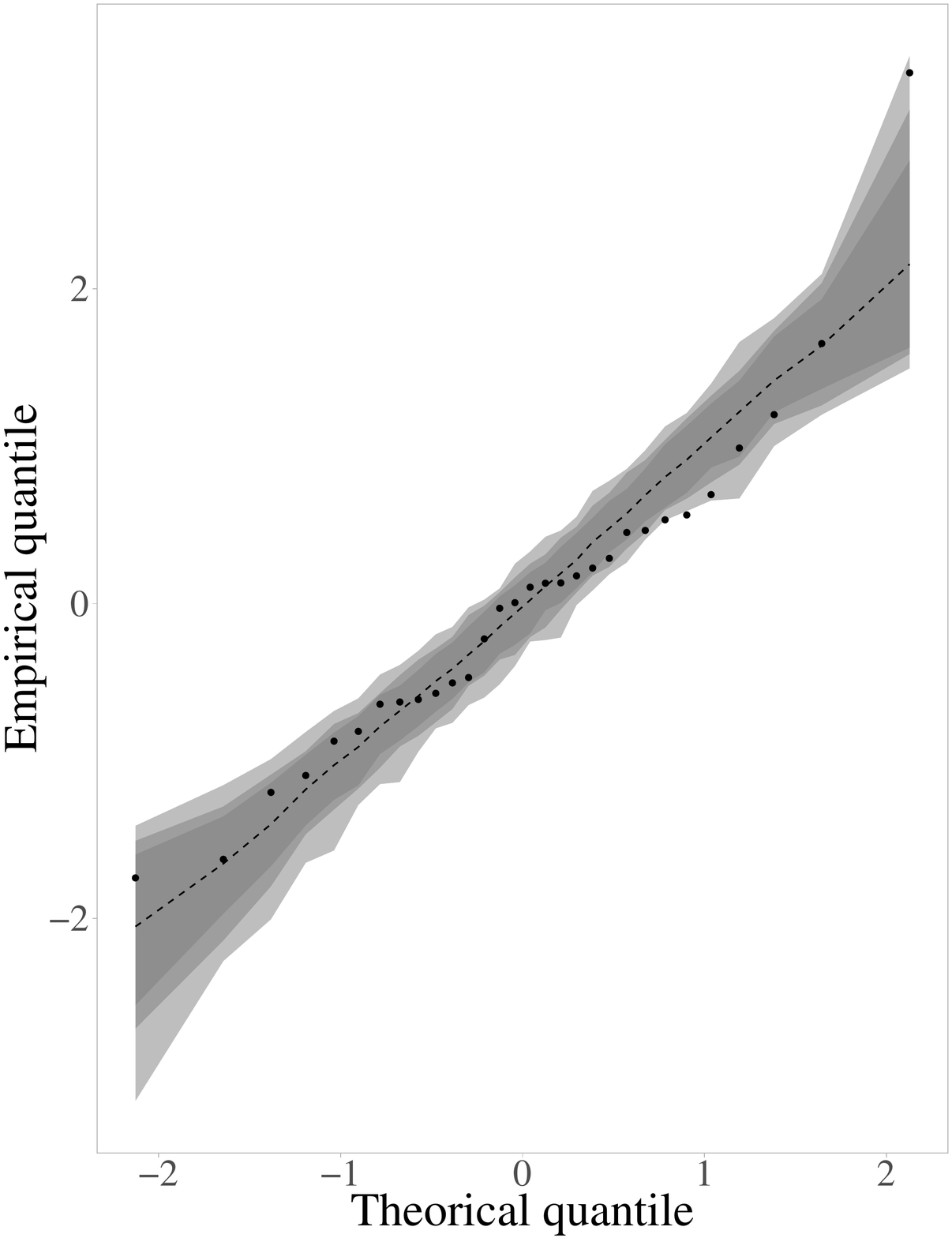}} \quad
\subfigure[][\textbf{GA}]{\includegraphics[height=5.5cm,width=5.5cm]{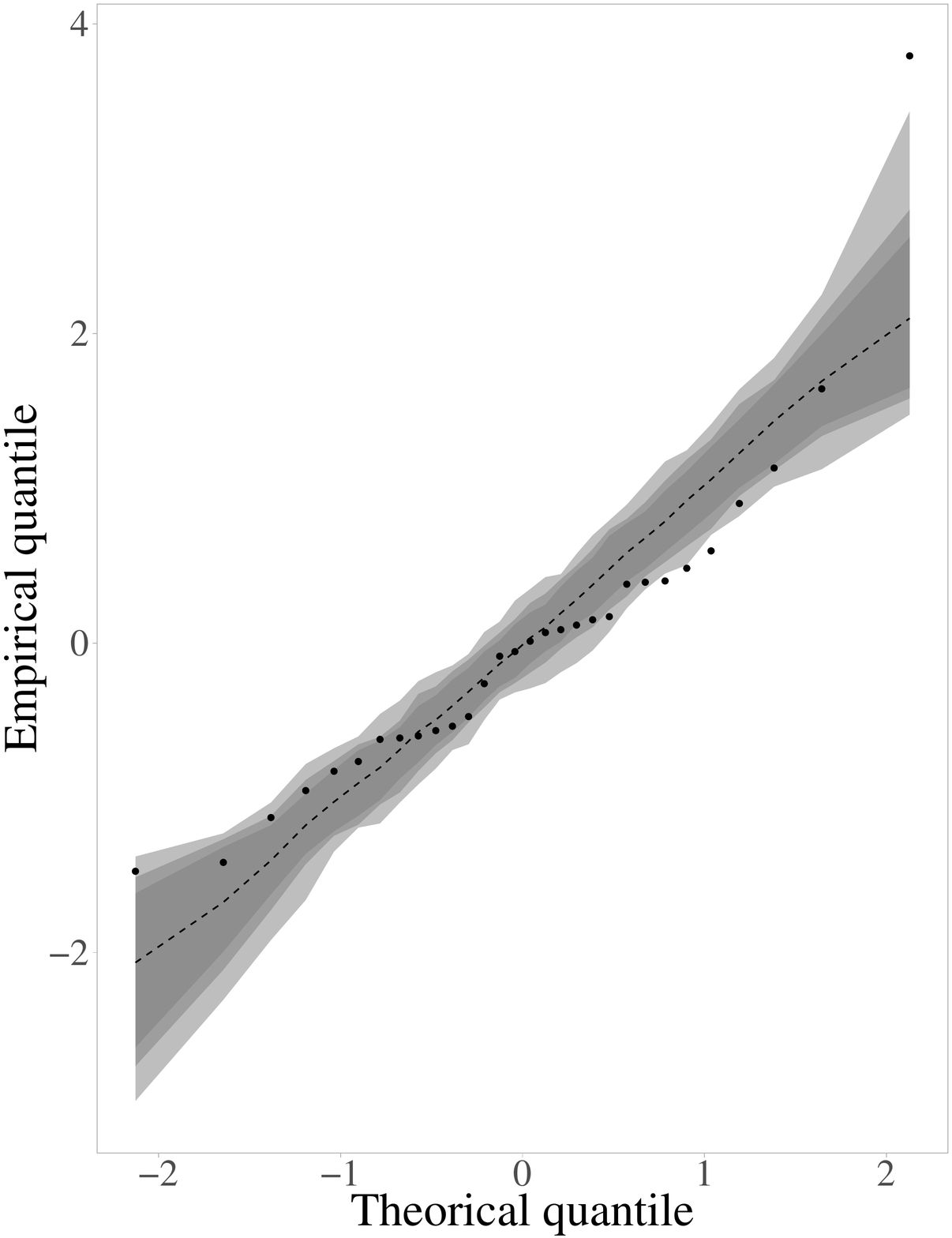}}\quad
\subfigure[][\textbf{RBS}]{\includegraphics[height=5.5cm,width=5.5cm]{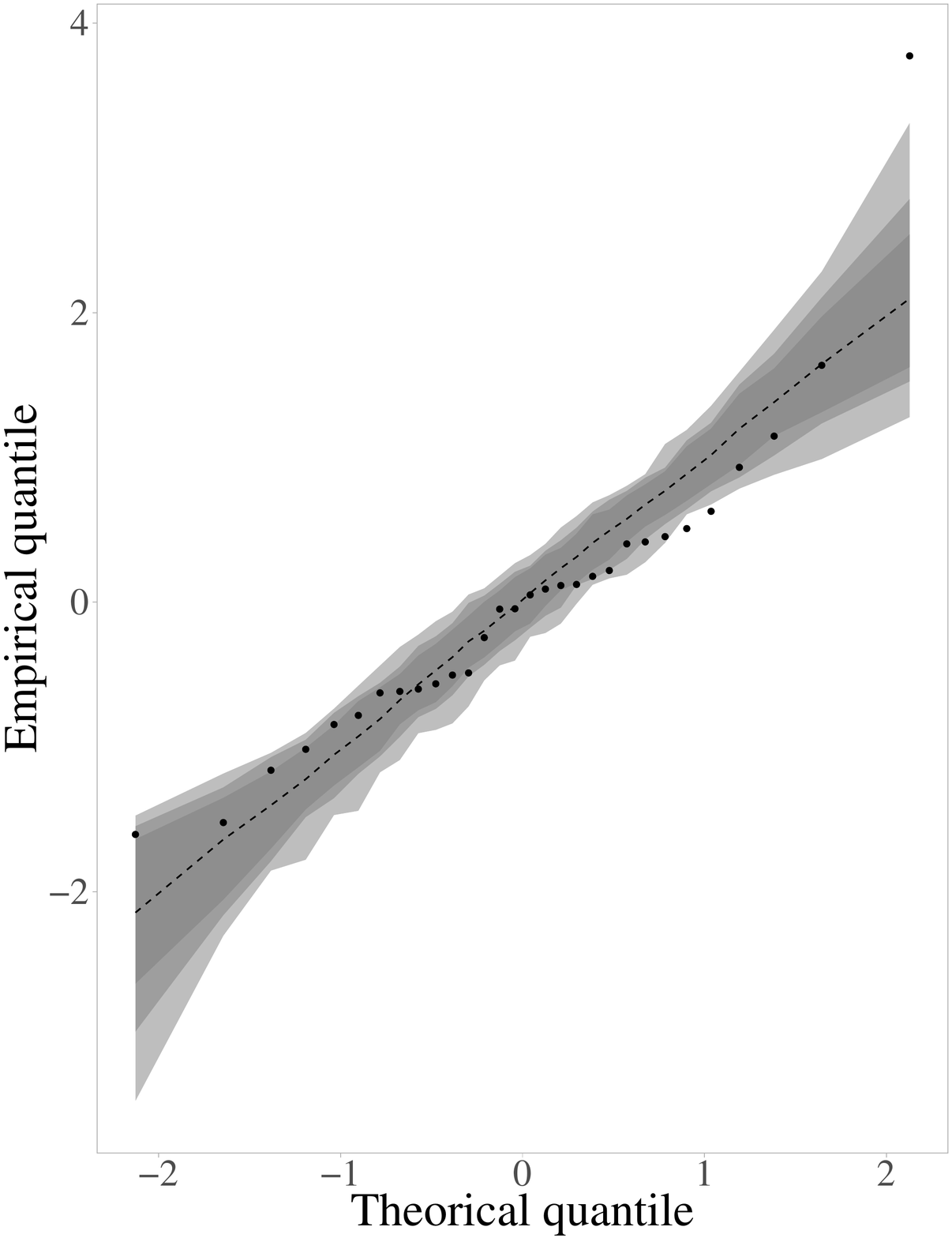}}
%\subfigure[][]{\includegraphics[height=3.5cm,width=5.5cm]{resbpmu1mu}} \quad
%\subfigure[][]{\includegraphics[height=3.5cm,width=5.5cm]{envelopebpmu1mu}}\quad
%\subfigure[][]{\includegraphics[height=3.5cm,width=5.5cm]{envelopebpmu1mu}}
%%\subfigure[$\phi = 100$]{\includegraphics[height=5.5cm,width=5.5cm]{density3}}
\caption{Simulated envelopes for the $r_i^\mathbf{Q}$ under perturbed BP (a), GA (b) and RBS (c) models.}
\label{fig:8}
\end{figure}

\begin{figure}[t]
\centering
\subfigure[][\textbf{BP}]{\includegraphics[height=5.5cm,width=5.5cm]{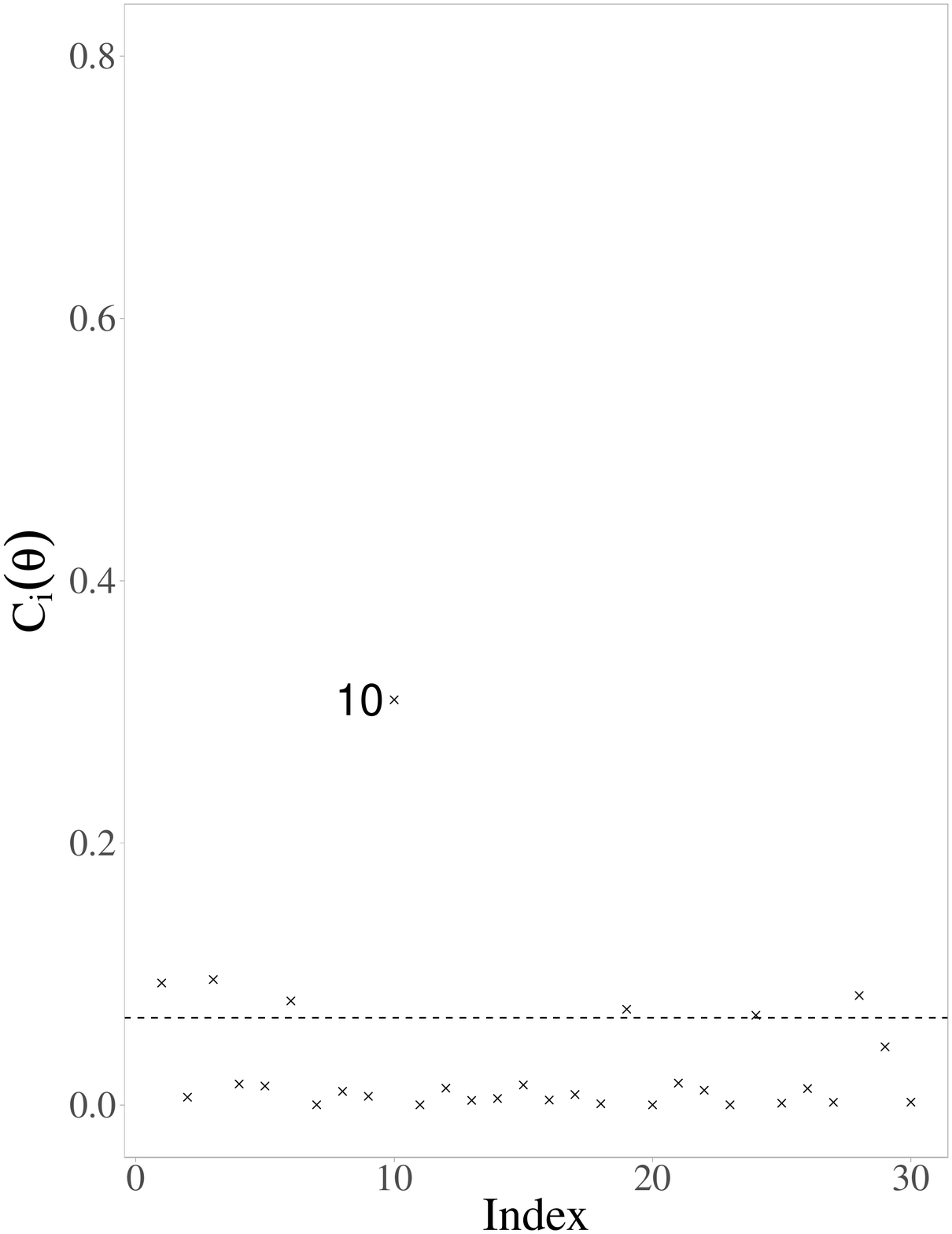}} \quad
\subfigure[][\textbf{GA}]{\includegraphics[height=5.5cm,width=5.5cm]{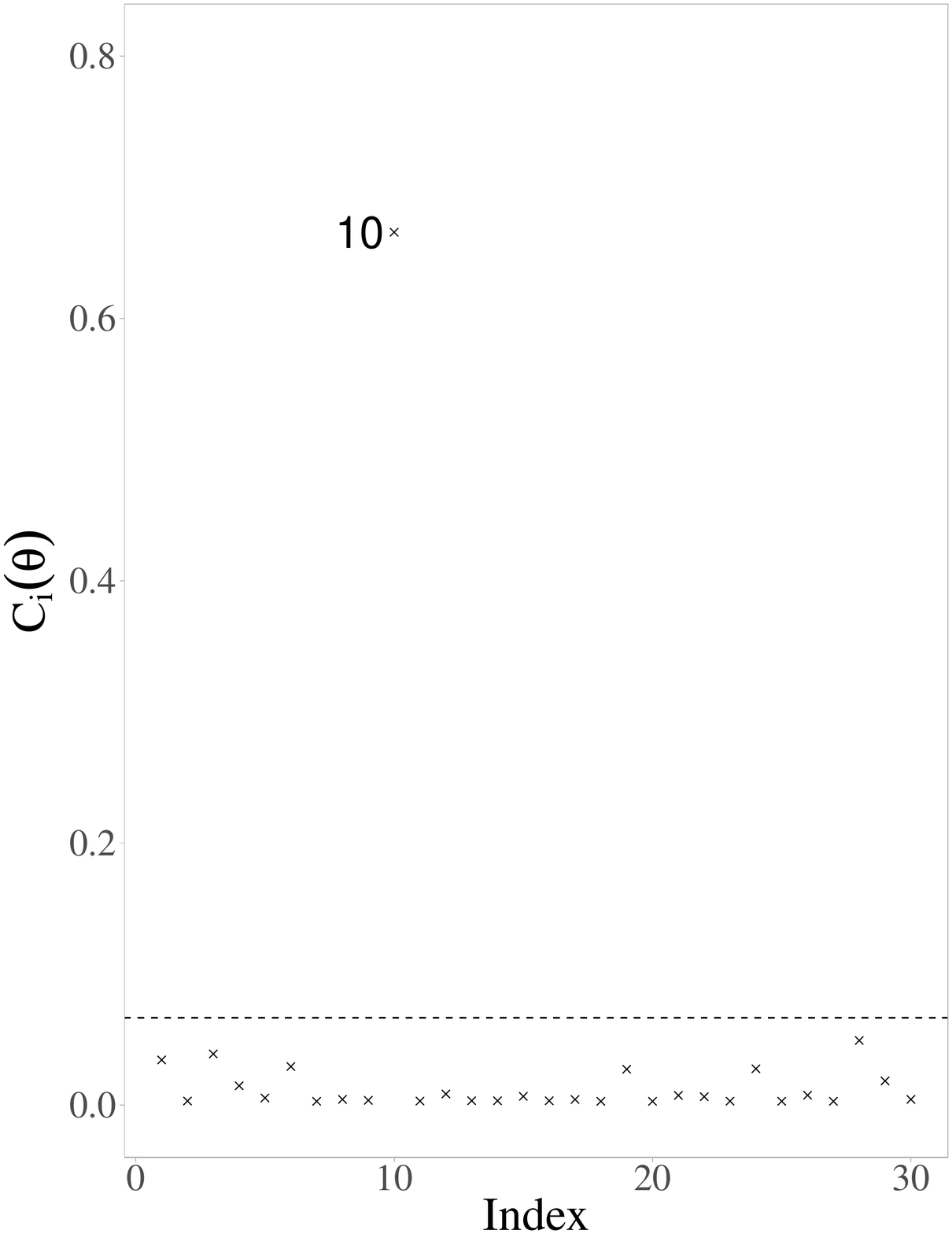}}\quad
\subfigure[][\textbf{RBS}]{\includegraphics[height=5.5cm,width=5.5cm]{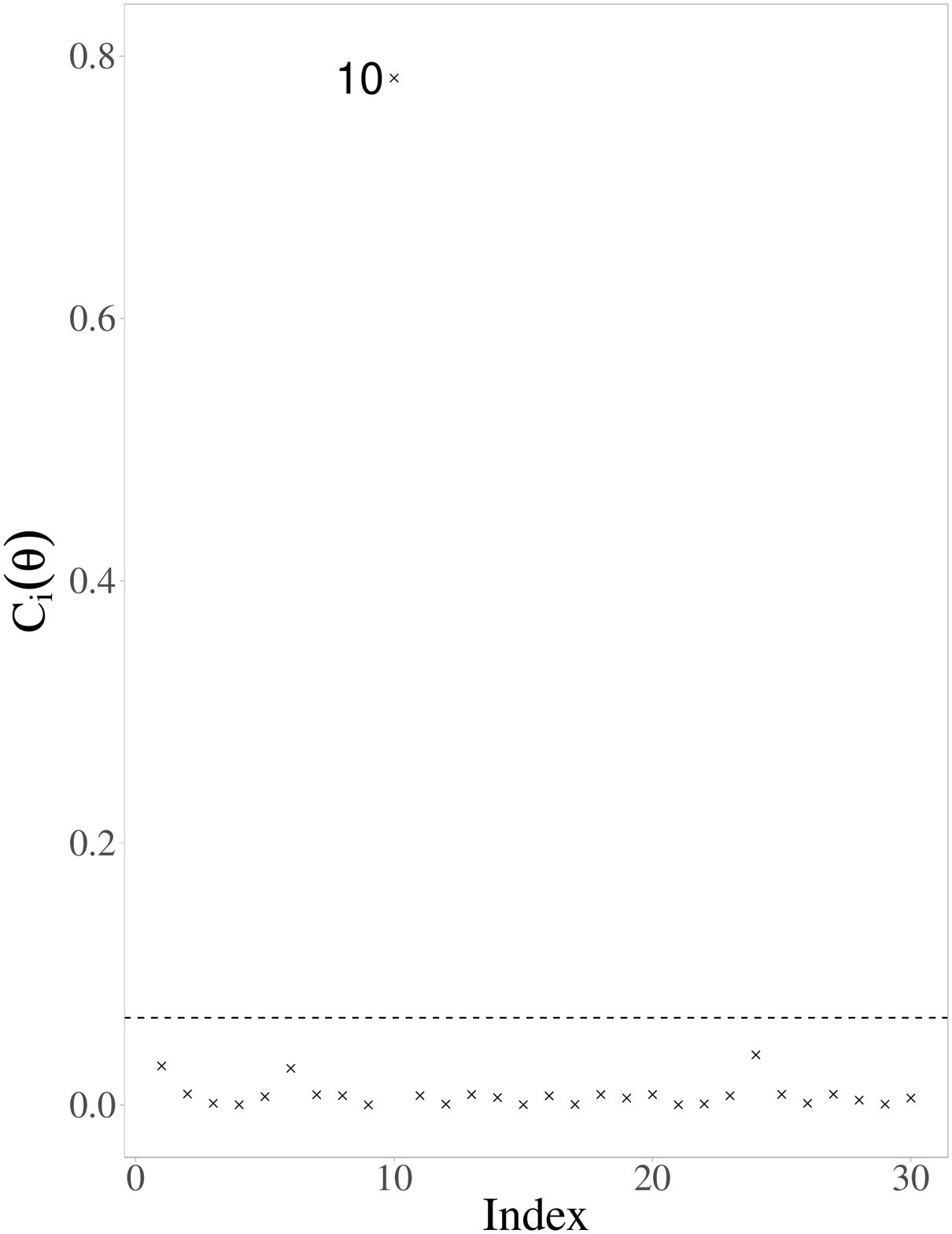}}
%\subfigure[][]{\includegraphics[height=3.5cm,width=5.5cm]{resbpmu1mu}} \quad
%\subfigure[][]{\includegraphics[height=3.5cm,width=5.5cm]{envelopebpmu1mu}}\quad
%\subfigure[][]{\includegraphics[height=3.5cm,width=5.5cm]{envelopebpmu1mu}}
%%\subfigure[$\phi = 100$]{\includegraphics[height=5.5cm,width=5.5cm]{density3}}
\caption{Index plot of  $C_i$ for $\widehat{\bm \theta}$, under perturbed BP (a), GA (b) and RBS (c) models.}
\label{fig:9}
\end{figure}

Now, we will introduce an additive perturbation on the 10$th$ observation (randomly selected) by making $y_{10} + 3 \, s_y$, where $s_y$ is the sample standard deviation of the response variable.
The purpose of this perturbation is to show that the BP model can be less sensitive to the presence of atypical observations in sampled data. Figures~\ref{fig:8}--\ref{fig:9}
present the fitted values against the residuals $r_i^\mathbf{Q}$, the simulated envelopes for the $r_i^\mathbf{Q}$ and the index plot of $C_i$ for $\widehat{{\bm \beta}}$ under the case-weights scheme, respectively, for the disturbed data.
From Figure \ref{fig:9}, we observe that that case 10 is the most influential for the GA and RBS models. From now on, therefore, we will follow our analyzes considering the BP regression model.
%
%
%\begin{figure}[t]
%\centering
%\subfigure[][\textbf{BP}]{\includegraphics[height=6.5cm,width=5.5cm]{resQ_BP}} \quad
%\subfigure[][\textbf{GA}]{\includegraphics[height=6.5cm,width=5.5cm]{resQ_GA}}\quad
%\subfigure[][\textbf{RBS}]{\includegraphics[height=6.5cm,width=5.5cm]{resQ_RBS}}
%%\subfigure[][]{\includegraphics[height=3.5cm,width=5.5cm]{resbpmu1mu}} \quad
%%\subfigure[][]{\includegraphics[height=3.5cm,width=5.5cm]{envelopebpmu1mu}}\quad
%%\subfigure[][]{\includegraphics[height=3.5cm,width=5.5cm]{envelopebpmu1mu}}
%%%\subfigure[$\phi = 100$]{\includegraphics[height=5.5cm,width=5.5cm]{density3}}
%\caption{Fitted values against the residuals $r_i^\mathbf{Q}$ under perturbed BP (a), GA (b) and RBS (c) models.}
%\label{fig:7}
%\end{figure}

\begin{figure}[t]
\centering
\subfigure[][\label{fig9a}]{\includegraphics[height=5.5cm,width=8.5cm]{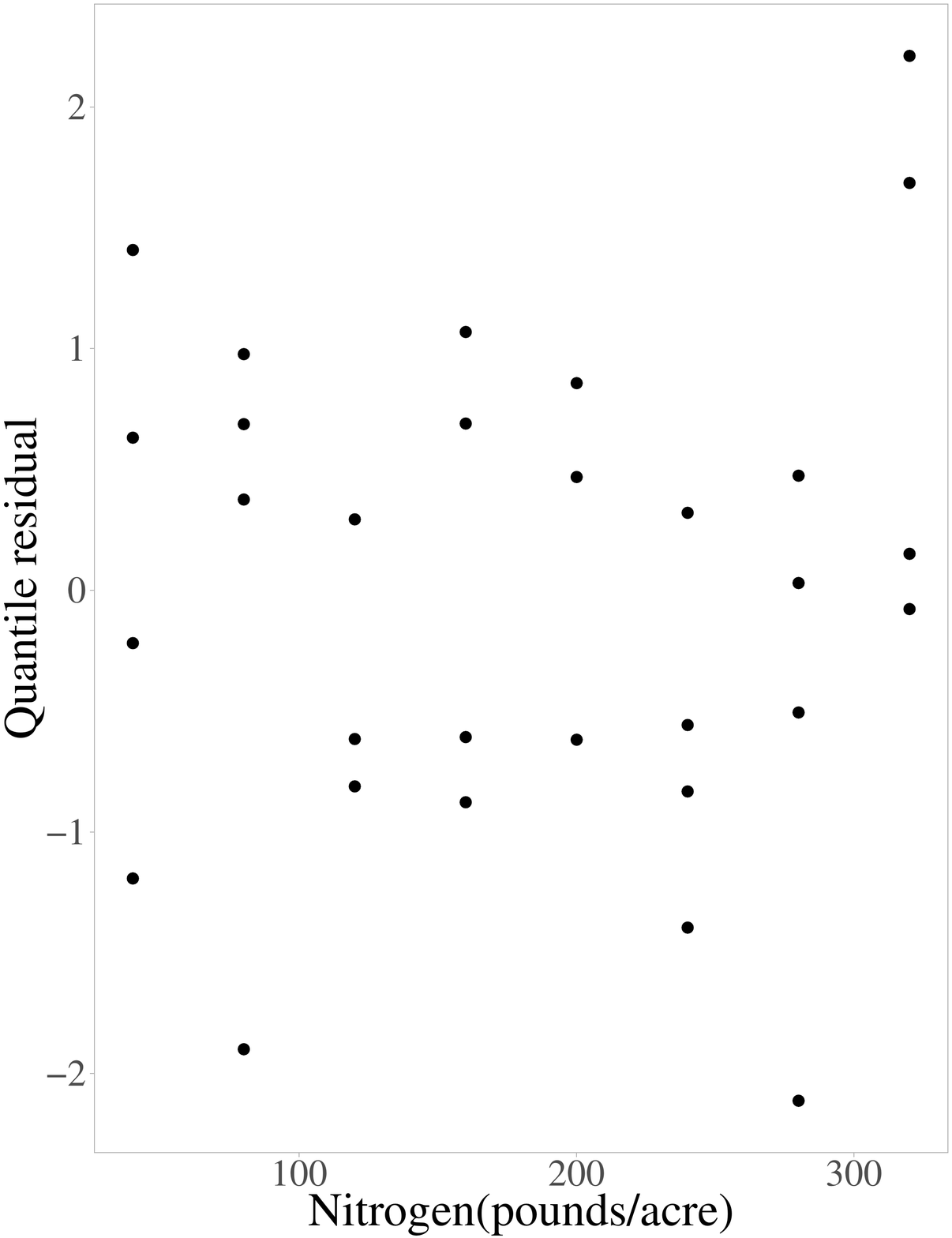}} \quad
\subfigure[][\label{fig9b}]{\includegraphics[height=5.5cm,width=8.5cm]{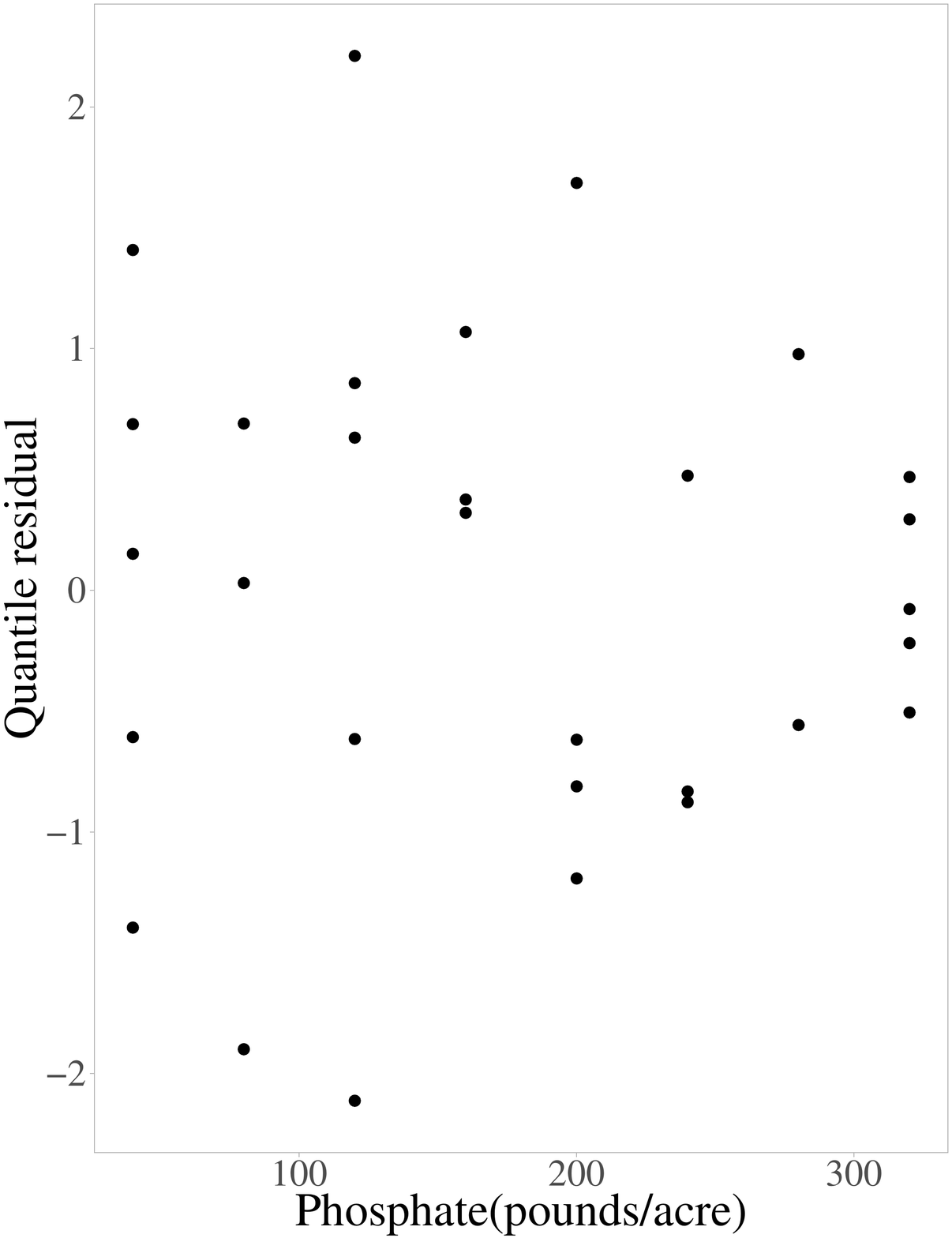}}
%\subfigure[$\phi = 100$]{\includegraphics[height=5.5cm,width=5.5cm]{density3}}
\caption{Nitrogen against the residuals $r_i^\mathbf{Q}$ (a) and phosphate against the residuals $r_i^\mathbf{Q}$ (b).}
\label{fig9}
\end{figure}

\begin{figure}[t]
\centering
\subfigure[][\label{fig10a}]{\includegraphics[height=5.5cm,width=8.5cm]{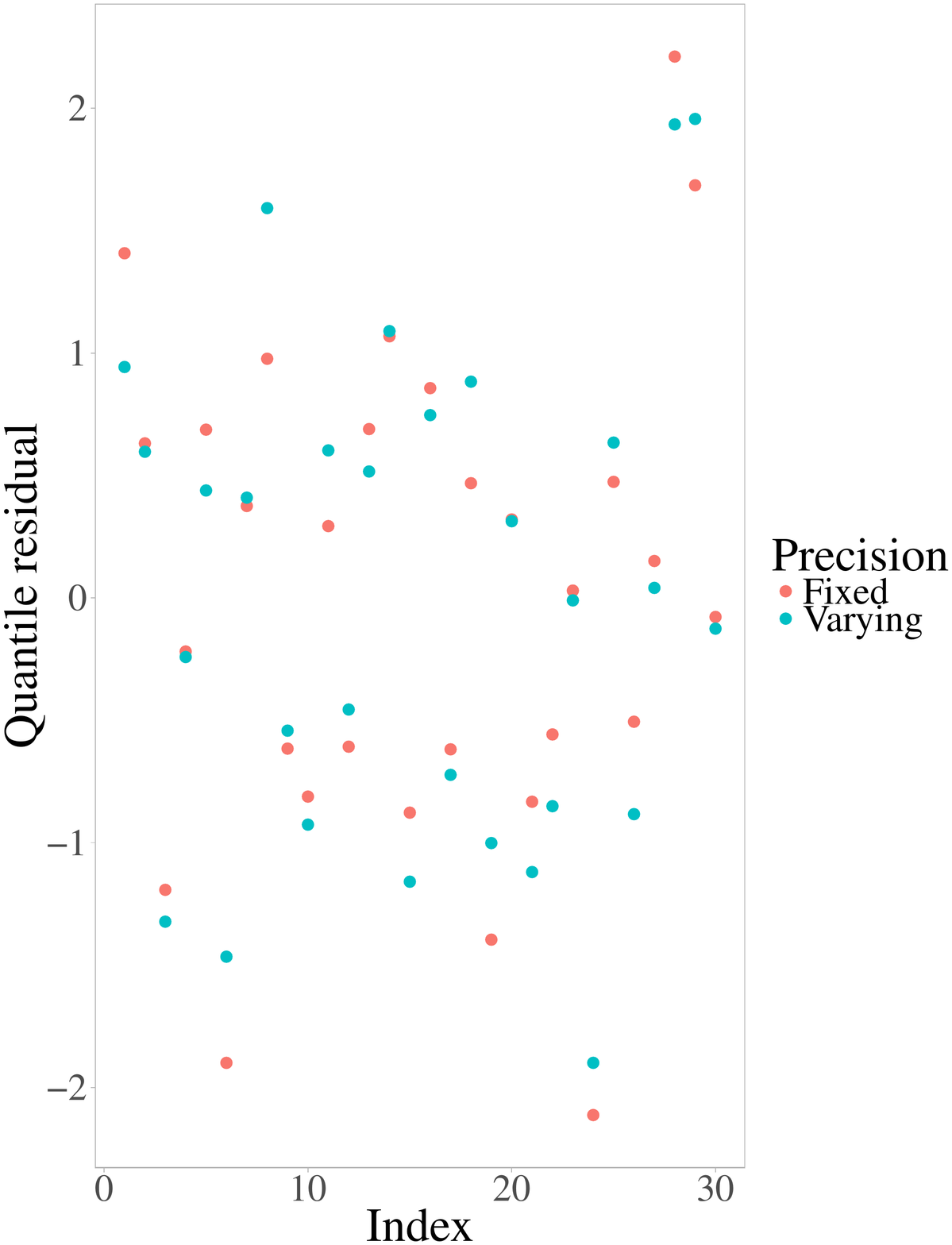}} \quad
\subfigure[][\label{fig10b}]{\includegraphics[height=5.5cm,width=8.5cm]{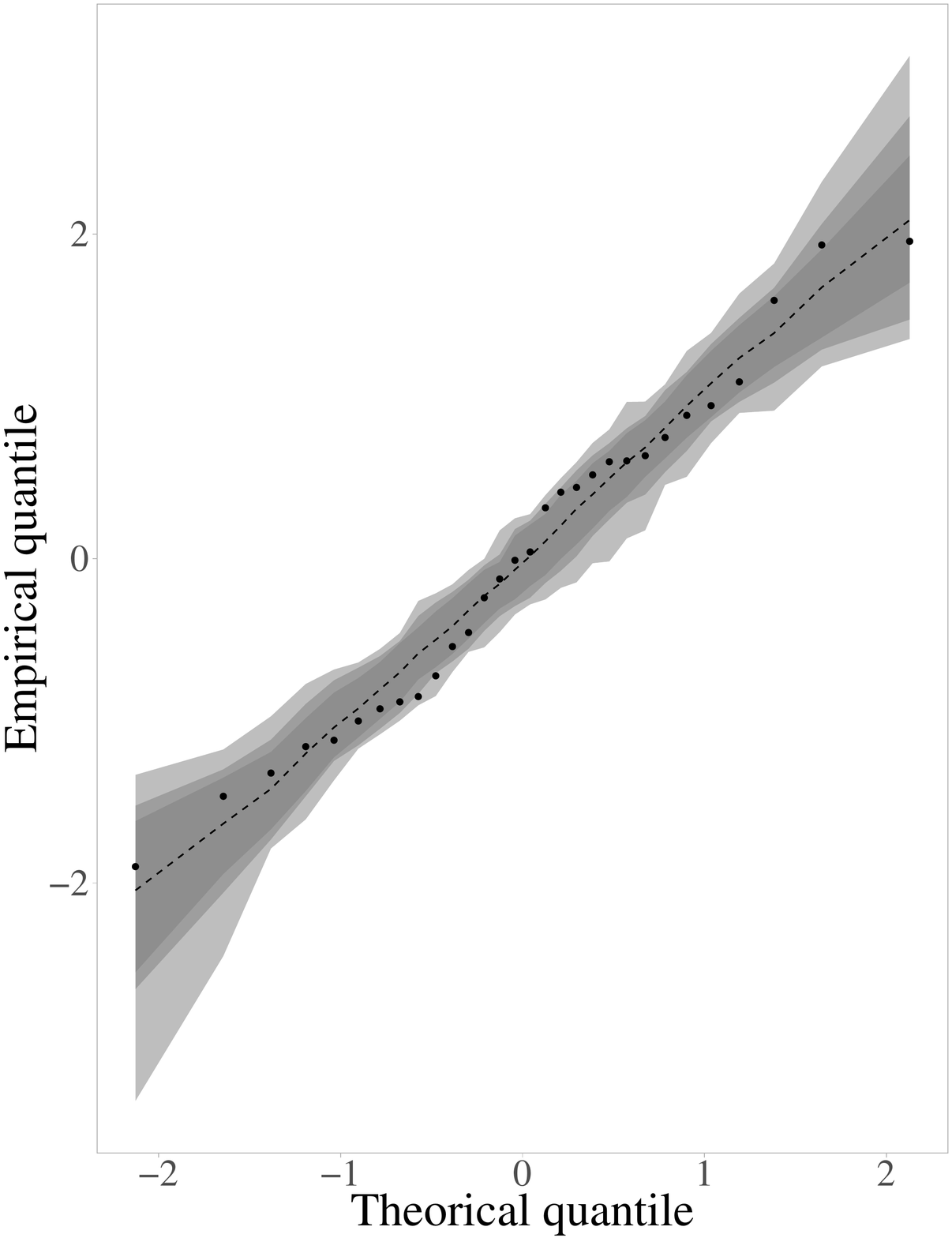}}
%%\subfigure[$\phi = 100$]{\includegraphics[height=5.5cm,width=5.5cm]{density3}}
\caption{Index plot of the residuals $r_i^\mathbf{Q}$(a) and simulated envelopes for the residuals $r_i^\mathbf{Q}$(b).}
\label{fig10}
\end{figure}

\begin{figure}[t]
\centering
\subfigure[][]{\includegraphics[height=5.5cm,width=5.5cm]{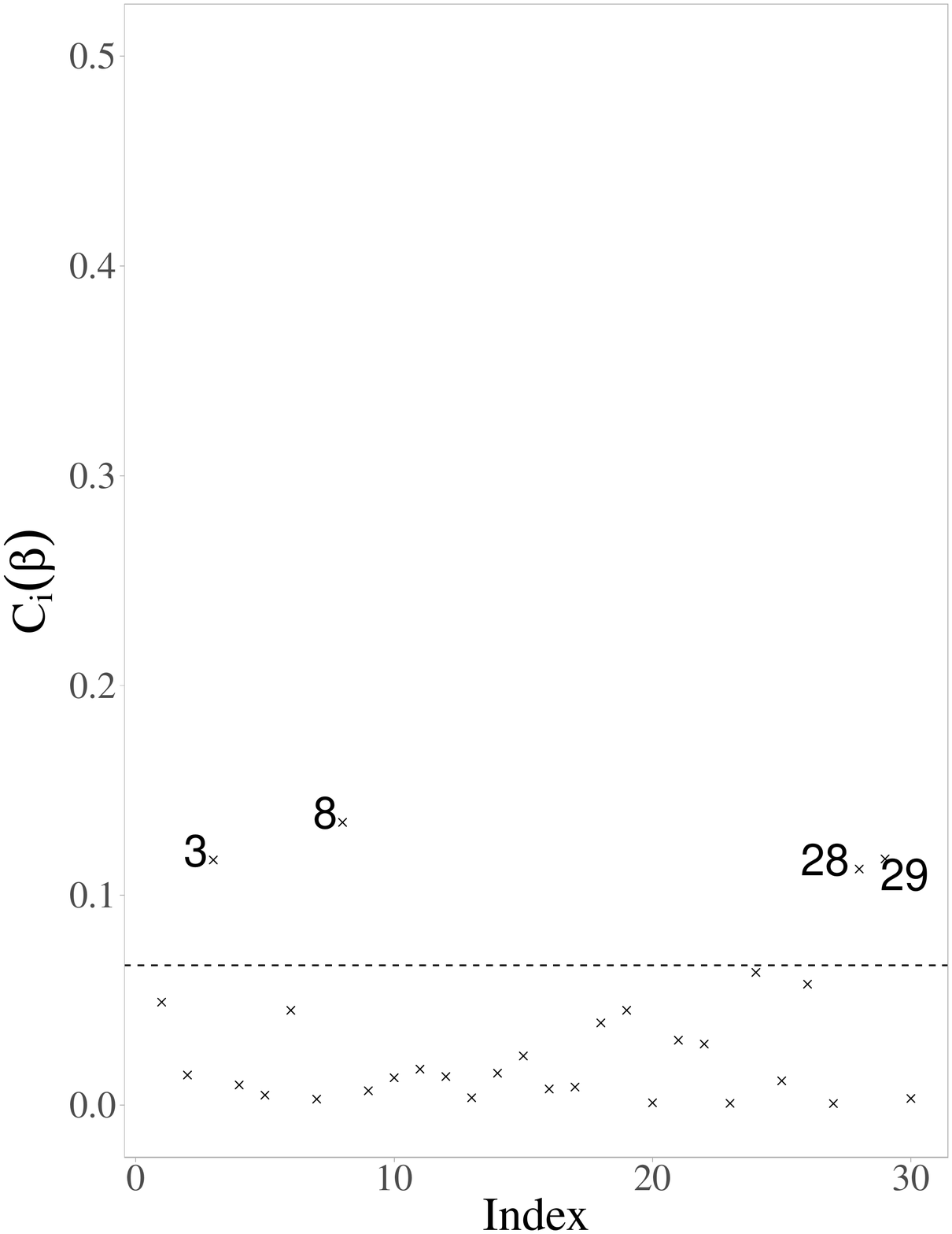}} \quad
\subfigure[][]{\includegraphics[height=5.5cm,width=5.5cm]{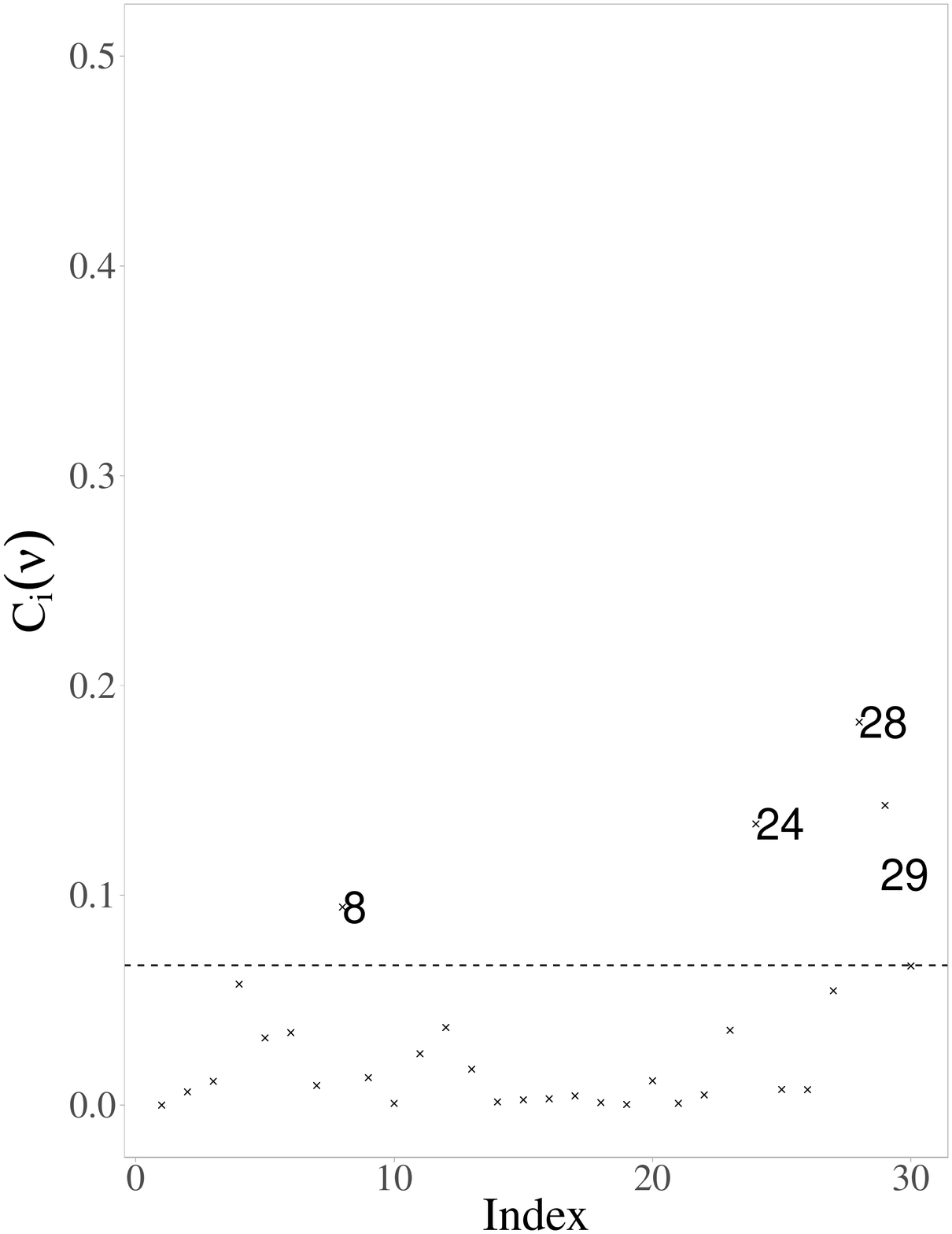}}\quad
\subfigure[][]{\includegraphics[height=5.5cm,width=5.5cm]{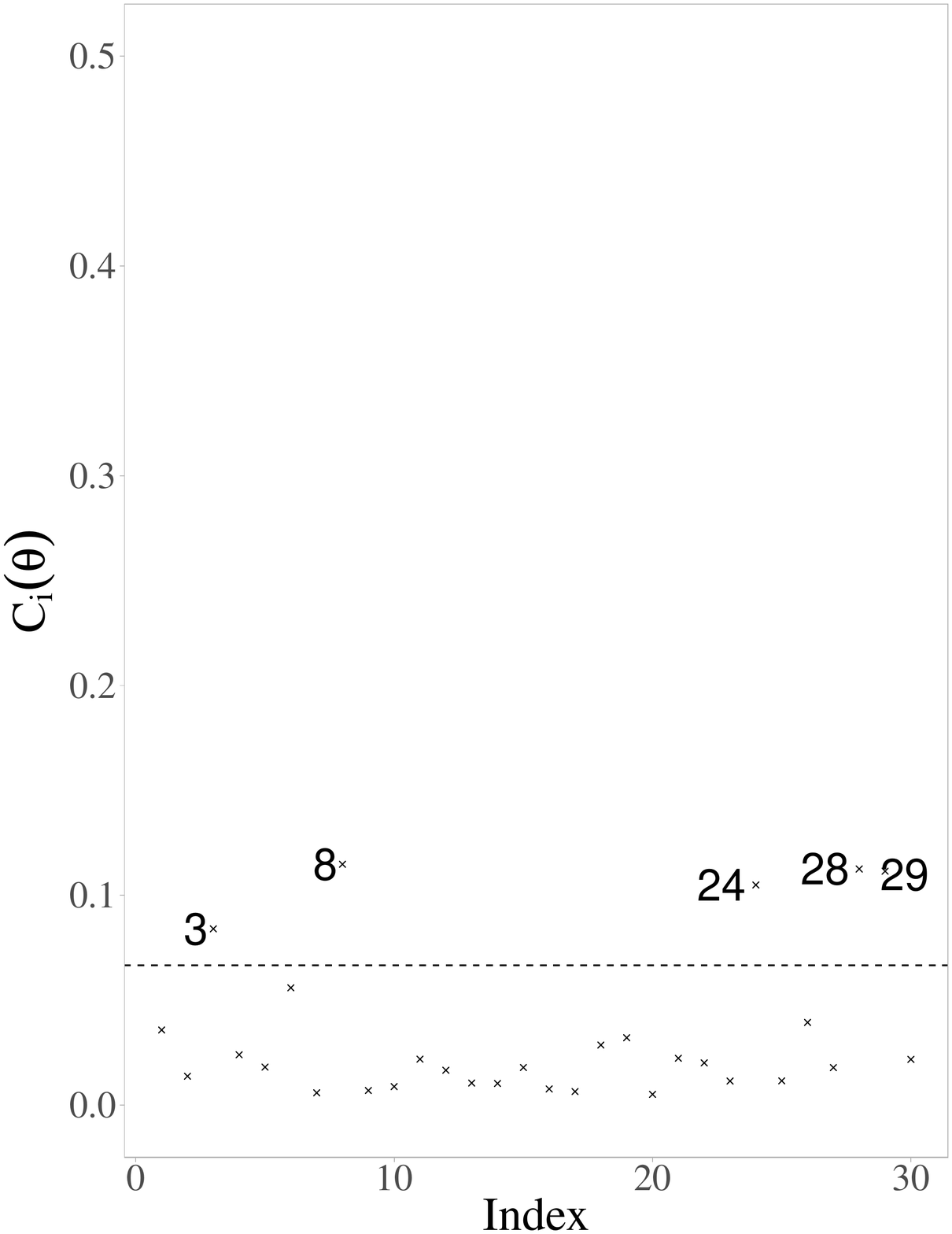}}
%\subfigure[][]{\includegraphics[height=3.5cm,width=5.5cm]{resbpmu1mu}} \quad
%\subfigure[][]{\includegraphics[height=3.5cm,width=5.5cm]{envelopebpmu1mu}}\quad
%\subfigure[][]{\includegraphics[height=3.5cm,width=5.5cm]{envelopebpmu1mu}}
%%\subfigure[$\phi = 100$]{\includegraphics[height=5.5cm,width=5.5cm]{density3}}
\caption{Index plot of  $C_i$ for $\widehat{\bm \beta}$ (left), $\widehat{\bm \nu}$ (center) and $\widehat{\bm \theta}$ (right) under the case-weights scheme.}
\label{fig11}
\end{figure}

\begin{figure}[t]
\centering
\subfigure[][]{\includegraphics[height=5.5cm,width=5.5cm]{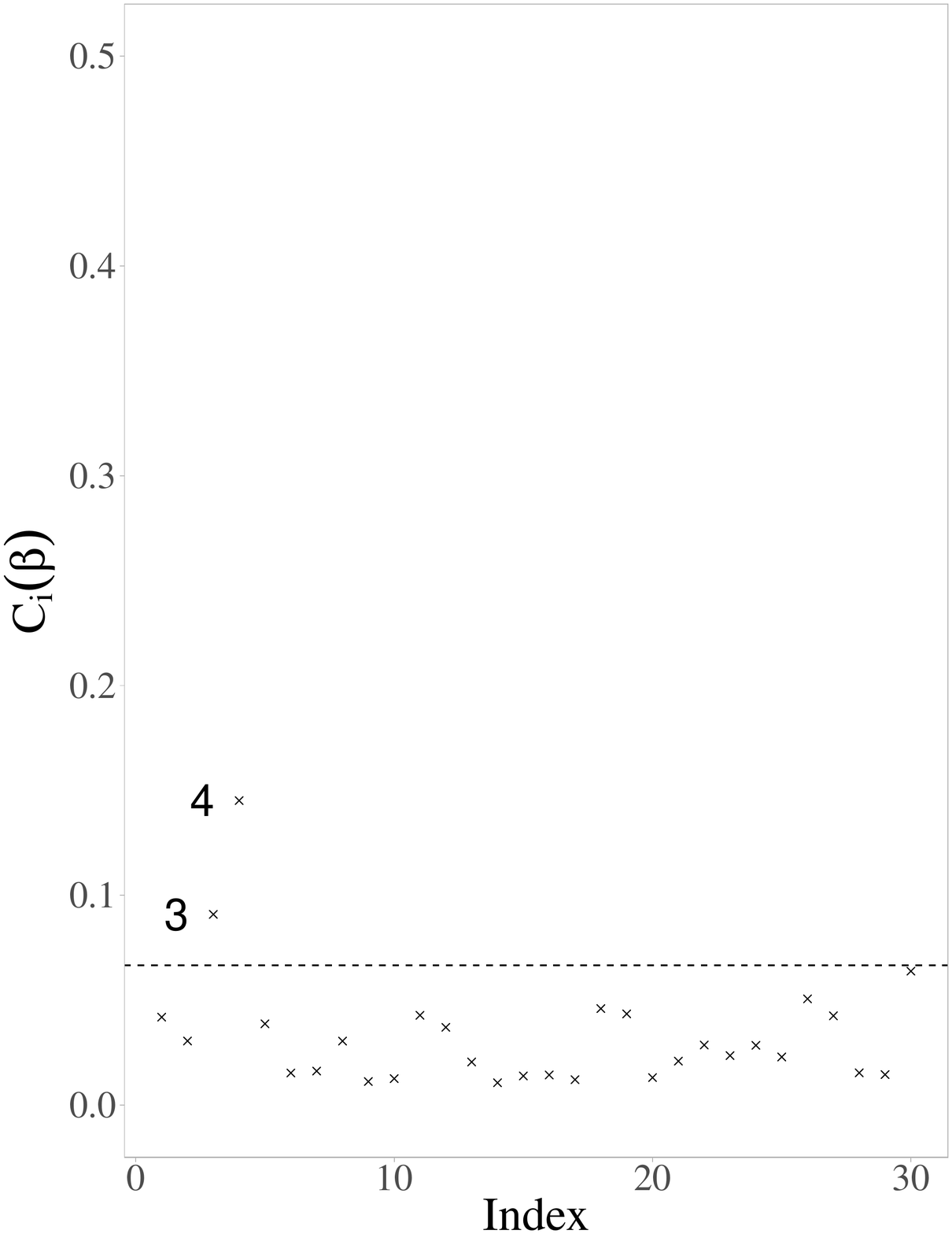}} \quad
\subfigure[][]{\includegraphics[height=5.5cm,width=5.5cm]{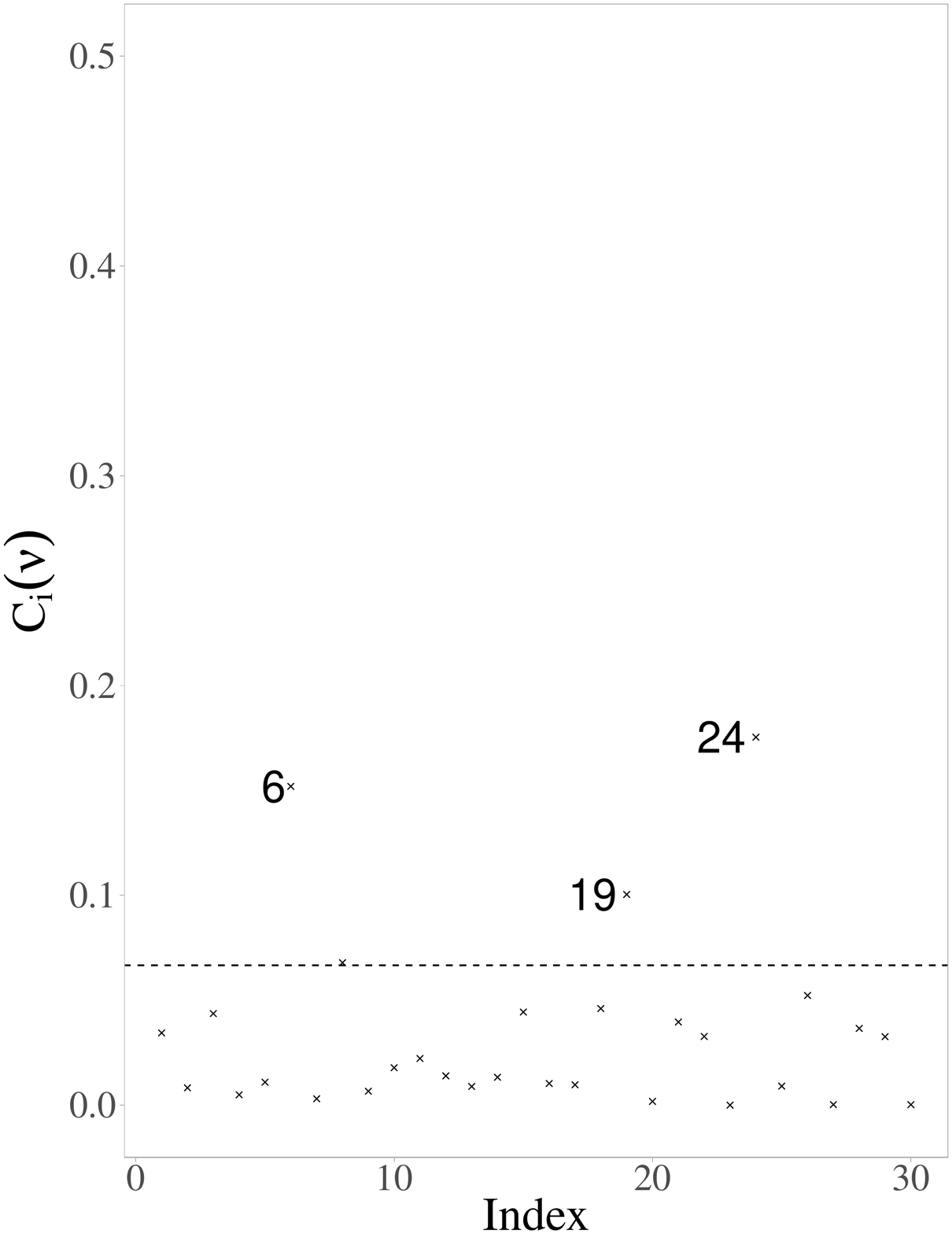}}\quad
\subfigure[][]{\includegraphics[height=5.5cm,width=5.5cm]{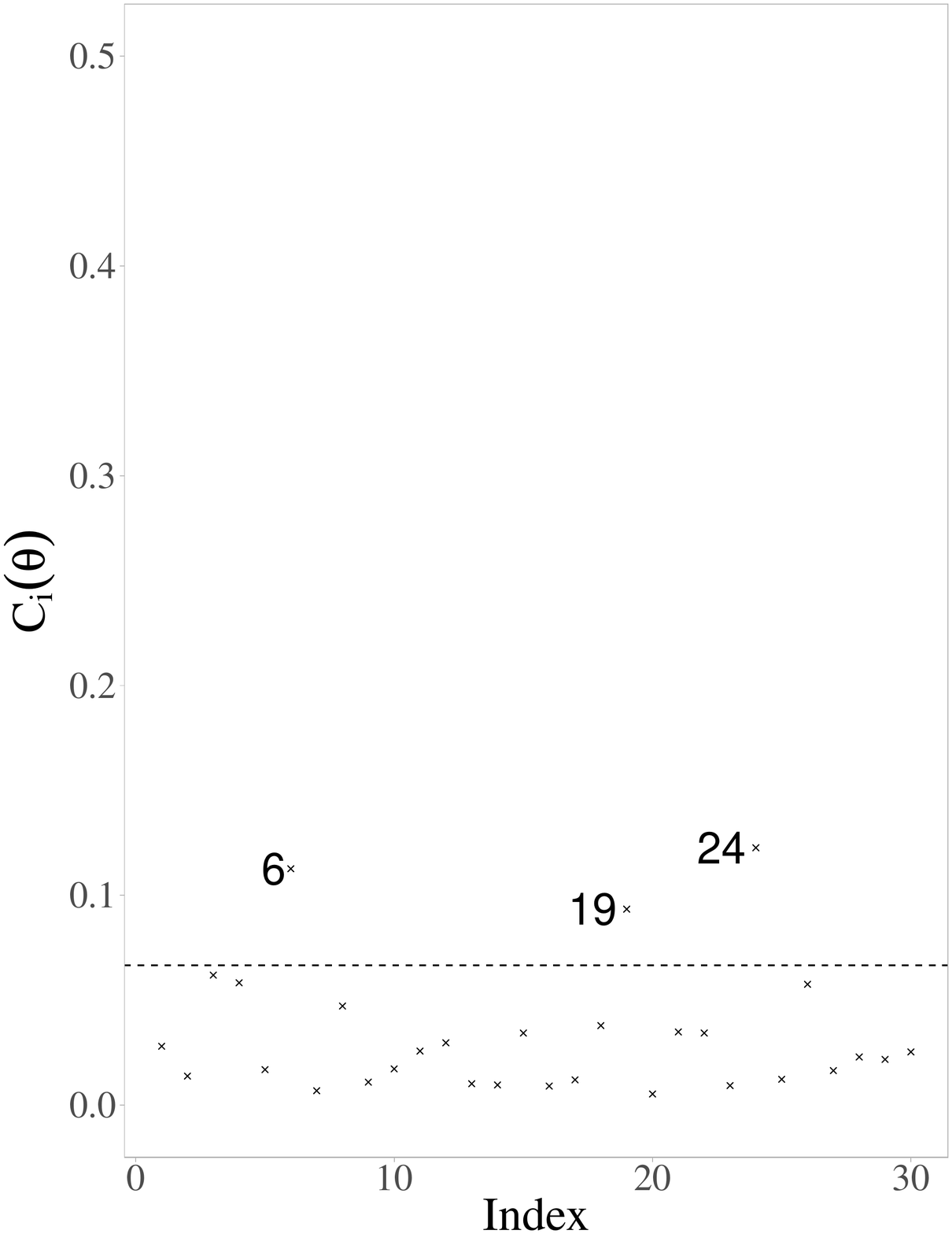}}
%\subfigure[][]{\includegraphics[height=3.5cm,width=5.5cm]{resbpmu1mu}} \quad
%\subfigure[][]{\includegraphics[height=3.5cm,width=5.5cm]{envelopebpmu1mu}}\quad
%\subfigure[][]{\includegraphics[height=3.5cm,width=5.5cm]{envelopebpmu1mu}}
%%\subfigure[$\phi = 100$]{\includegraphics[height=5.5cm,width=5.5cm]{density3}}
\caption{Index plot of  $C_i$ for $\widehat{\bm \beta}$ (left), $\widehat{\bm \nu}$ (center) and $\widehat{\bm \theta}$ (right) under the response perturbation scheme.}
\label{fig12}
\end{figure}

%
%
%%\begin{table}[H]
%%\centering
%%\caption{Model selection criteria.}\label{aplic2}
%%\renewcommand{\arraystretch}{1.3}
%%%\resizebox{\linewidth}{!}{
%%\begin{tabular}{lccc}
%%\hline
%%Measure&BP&Gamma&IG\\  \hline
%%log-likelihood & $-$112.19  & $-$112.30  & $-$113.86   \\
%%AIC  &232.38  & 232.59  &  235.71 \\
%%BIC   &237.99  & 238.20  & 241.32  \\
%%\hline
%%\end{tabular}
%%%}
%%\end{table}
%
%Figure~\ref{fig7} presents plots of quantile residuals against fitted values and
%simulated envelopes for the BP model. Notice that most residuals lie inside the intervals and that there is no noticeable pattern in the residuals. Visual inspection this plots reveals that all residuals are inside the envelope, thus indicating that the
%distributional assumptions.
%
%\begin{figure}[t]
%\centering
%\subfigure[][\label{fig7a}]{\includegraphics[height=6.5cm,width=8.5cm]{resbpmu1mu}} \quad
%\subfigure[][\label{fig7b}]{\includegraphics[height=6.5cm,width=8.5cm]{envelopebpmu1mu}}
%%\subfigure[$\phi = 100$]{\includegraphics[height=5.5cm,width=5.5cm]{density3}}
%\caption{Fitted values against the residuals $r_i^\mathbf{Q}$(a) and simulated envelopes for the residuals $r_i^\mathbf{Q}$(b).}
%\label{fig7}
%\end{figure}
%
Non-constant variance in can be diagnosed by residual plots. Figure~\ref{fig9b} note that the residual plot shows a pattern that indicates an evidence of a non-constant precision because the variability is higher for lower phosphate concentrations. Thus, we will consider the following model for precision of the BP regression model
\[
\log(\phi_i) = \nu_0 + \nu_1\, x_{2i}, \quad i = 1, \ldots, 30.
\]
The ML estimates of its parameters, with estimated asymptotic standard errors (SE) in parenthesis, are: $\widehat\beta_0 = 0.5207 \, (0.2788)$, $\widehat\beta_1 = 0.3506 \, (0.0330)$, $\widehat\beta_2 = 0.3990 \, (0.0423)$, $\widehat\nu_0 = 2.7027 \, (0.6650)$ and $\widehat\nu_1 = 0.0072 \, (0.0033)$.  Note that the coefficients are statistically significant at
the usual nominal levels. We also note that there is a positive relationship
between the mean response (the productivity of corn) and nitrogen, and that there is a positive relationship between the mean response
and the phosphate. Moreover, the likelihood ratio test for varying precision (given in Equation \eqref{lr}) is significant at the level of $5\%$ ($p$-value =0.0412), for the BP regression model with the structure above.
In Figure~\ref{fig10a}, we note that the quantile residuals under the BP regression model with varying precision have a smaller amplitude of their values. In Figure~\ref{fig10b}, we can see that there is no noticeable pattern in the residuals and  there is no evidence against the distributional assumptions.
In Figures~\ref{fig11}-\ref{fig12}, we present diagnostics for the BP regression model with varying precision. We can see in these figures that no observations stood out as potentially influential.
%
%
%
%%To evaluate the impact of the removed observations, we will define the relative chance that is given by%%\[
%%\mathbb{RC} =\Big|\frac{\hat \theta_i - \hat \theta_{i(j)}}{\hat \theta_i}\Big|, \quad i=1,\ldots, \#\bm\theta, \quad j \in \mathcal{P}.
%%\]
%%%
%%where $\#A$ represents the cardinality of the set $A$ and $\mathcal{P}$ is the power set of the set of all potentially influential observations. Table~XX we present the corresponding $\mathbb{RC}$ and inferential changes are not detected when the observations are removed.
%%

\section{Concluding remarks}\label{sec:8}

In this paper, we have developed a new parameterized
BP distribution in terms of the mean and precision parameters. The variance function of the proposed model assumes a quadratic form.
Furthermore, we have proposed a new regression model for modelling
asymmetric positive real data. 
%We use the new parametrization in which a function of the mean of the response is given by a linear predictor that is defined by regression parameters and explanatory variables. 
The proposed parameterization allows for a precision
parameter, which also has a systematic component. An advantage of the proposed BP regression model in relation to the GA and RBS regression models is its flexibility for working with positive real data with high skewness,
i.e., the proposed model may serve as a good alternative to the GA and RBS regression
models for modelling asymmetric positive real data.

Maximum likelihood inference is implemented for estimating the
model parameters and its good performance has been evaluated by
means of Monte Carlo simulations. Furthermore, we provide closed-form expressions for the score function and for Fisher's information matrix.
Diagnostic tools have been obtained
to detect locally influential data in the maximum likelihood estimates.
In particular, appropriate matrices for assessing local influence on the parameter estimates under different perturbation
schemes are obtained. We have proposed two types of residuals for the proposed model and conducted a simulation study to establish their
empirical properties in order to evaluate their performances.
Finally, an application using a real data set was presented and discussed.

As part of future work, there are several extensions of the new model not considered in
this paper that can be addressed in future research:
an extension of BP regression model that allow for covariates to be measured
with error, zero-augmented BP regression assumes that the response variable
has a mixed continuous-discrete distribution with probability mass at zero,
an extension of BP regression model with censored data may be developed.
Additionally, future work should explore other estimation methods for the BP
regression model as, for instance, the Bayesian, improved estimation \citep{CoxSnell86} and EM-algorithm approaches.
Finally, one may develop an \textbf{R} package to perform inference in the BP regression model.
Work on these problems is currently
under progress and we hope to report these findings in a future paper. We hope that this new model may attract wider applications
in regression analysis.

%\section*{Acknowledgments}
%We thank the associate editor and the reviewer for the constructive comments.

\section*{Appendix}
\appendix
\section{Score vector}
\label{appA}

First we will show how the elements of the score vector for this class of models were obtained. This elements are
obtained by differentiating the log-likelihood function, given in \eqref{logm}, with respect to the $\beta_j$ and $\nu_r$ and are given by

\begin{eqnarray*}% \label{eq:sc}
u_{\beta_j} &=& \pdv{\ell(\bm{\beta}, \bm{\nu})}{\beta_j}  = \sum_{i=1}^n \dot{d}^{[i]}_{\mu} \, a_i \, x_{ij}, \quad j = 1, \ldots, p, \nonumber \\
u_{\nu_r} &=&  \pdv{\ell(\bm{\beta}, \bm{\nu})}{\nu_r} = \sum_{i=1}^n \dot{d}^{[i]}_\phi\, b_i\, z_{ir}, \quad r = 1, \ldots, q,
\end{eqnarray*}
where $a_i = \pdv{\mu_i}{\eta_{1i}}$, $b_i = \pdv{\phi_i}{\eta_{2i}}$ and from~\eqref{logm} we have that note that $\partial \ell_i(\mu_i, \phi_i)/\partial \mu_i $ and $\partial \ell_i(\mu_i, \phi_i)/\partial \phi_i$, $i = 1, \ldots, n$, are given by
\begin{eqnarray}\label{deriv1}
\dot{d}^{[i]}_{\mu}=\pdv{\ell(\mu_i, \phi_i)}{\mu_i} &=&  (1+\phi_i)\left\{\log\left(\frac{y_i}{1+y_i}\right) -\left[\Psi^{(0)}(\mu_i(1+\phi_i) ) -\Psi^{(0)}(\mu_i(1+ \phi_i) + \phi_i +2)\right]\right\},\nonumber\\
&=& (1+\phi_i)(y_i^{*} - \mu_i^{*}),
\end{eqnarray}
and
\begin{eqnarray}\label{deriv2}
\dot{d}^{[i]}_{\phi} = \pdv{\ell(\mu_i, \phi_i)}{\phi_i}&=&\mu_i\left[\log(y_i) - \Psi^{(0)}(\mu_i(1+\phi_i))\right] -  (1+\mu_i)\left[\log(1+y_i) - \Psi^{(0)}(\mu_i(1+\phi_i) + \phi_i +2)\right] \nonumber \\
&&-\Psi^{(0)}(\phi_i+2),\nonumber \\
 &=& \log\left(\frac{y_i^{\mu_i}}{(1+y_i)^{1+\mu_i}} \right) - \mu_i\,\Psi^{(0)}(\mu_i(1+\phi_i)) + (1+\mu_i)\Psi^{(0)}(\mu_i(1+\phi_i) + \phi_i +2)\nonumber \\
 &&- \Psi^{(0)}(\phi_i+2),\nonumber\\
 &=& y^\star_i - \mu_i\,\left(\mu_i^{*} - \frac{\gamma_i}{\mu_i}\right) = y^\star_i - \mu^\star_i,
\end{eqnarray}
where $y_i^{*}=\log\left(\frac{y_i}{1+y_i}\right)$, $\mu_i^{*} = \left[\Psi^{(0)}(\mu_i(1+\phi_i) ) -\Psi^{(0)}(\mu_i(1+ \phi_i) + \phi_i +2)\right]$, $y^\star_i = \log\left(\frac{y_i^{\mu_i}}{(1+y_i)^{1+\mu_i}} \right)$ and $\mu^\star_i = \mu_i\,\left(\mu_i^{*} - \frac{\gamma_i}{\mu_i}\right)$ with $\gamma_i = \Psi^{(0)}(\mu_i(1+\phi_i) + \phi_i +2) - \Psi^{(0)}(\phi_i+2)$.

The score function can be written in matrix form as
\begin{eqnarray*}%\label{scv}
\mathbf{U}_{\bm{\theta}} = \begin{bmatrix}
 \mathbf{X}^\top \, \bm{\Phi} \, \mathbf{D}_1\, (\mathbf{y}^{*}-\bm{\mu^*})    \\
 \mathbf{Z}^\top\,\mathbf{D}_2\,(\mathbf{y}^{\star}-\bm{\mu^\star})
\end{bmatrix},
\end{eqnarray*}
where $\mathbf{D}_1 = \left[a_i\, \delta_{ij}\right]$, $\mathbf{D}_2 = \left[b_i\, \delta_{ij}\right]$, $\bm{\Phi} = \left[(1+\phi_i)\,\delta_{ij}\right]$,
$\mathbf{y}^{*} = (y_1^{*}, \ldots, y_n^{*})\top$, $\mathbf{y}^{\star} = (y_1^{\star}, \ldots, y_n^{\star})\top$, $\bm{\mu}^{*} = (\mu_1^{*}, \ldots, \mu_n^{*})^\top$, $\bm{\mu}^\star = (\mu^\star_1, \ldots, \mu^\star_n)^\top$ and $\delta_{ij}$ is the Kronecker delta for $i,j=1, 2, \ldots, n$.

\section{Hessian matrix and Fisher's information matrix}
\label{appB}

Under the BP regression model defined in Section~, the second
derivative of $\ell(\bm{\beta},\bm \nu)$ with respect:

\vspace{0.5cm}
\noindent
 \textbf{(i)} to $\beta_j$ and $\beta_l$, for $j,l = 1, \ldots,p$, is
\begin{eqnarray*}
\pdv[2]{\ell(\bm{\beta},\bm \nu)}{\beta_j}{\beta_l}
&=&
\sum_{i=1}^{n}\underbrace{\left\{ \ddot{d}^{[i]}_{\mu_i}\, a_i^2  + \dot{d}^{[i]}_{\mu_i}\, a'_i \, a_i \right\} }_{c_i}
x_{ij}x_{il},
 \label{c}
\end{eqnarray*}
where $a'_i = \pdv{a_i}{\mu_i}$, $\dot{d}^{[i]}_{\mu_i}$ is defined in \eqref{deriv1} and
\begin{eqnarray*}\label{deriv3}
\ddot{d}^{[i]}_{\mu} = \pdv[2]{\ell(\mu_i, \phi_i)}{\mu_i} &=&  -(1+\phi_i)^2\left[\Psi^{(1)}(\mu_i(1+\phi_i) ) - \Psi^{(1)}(\mu_i(1+ \phi_i) + \phi_i +2)\right],
\end{eqnarray*}
where $\Psi^{(1)}(z) = \frac{d}{d z}\Psi^{(0)}(z)$.

%\[
%\ddot{d}^{[i]}_{\mu_i}  = y_i^\circ - \mu^\circ_i
%\]
%with \( \pdv{\varphi_i}{\mu_i}=  \phi \left\{\frac{ [1+2\mu_i]\textrm{V}(\mu_i) - \textrm{V'}(\mu_i)[\mu(1+\mu_i)]   }{\textrm{V}(\mu_i)^2} \right\}\), \(\pdv[2]{\varphi_i}{\mu_i} = \phi \left\{\frac{ 2\textrm{V}(\mu)^2 - \textrm{V}(\mu)\textrm{V''}(\mu)[\mu(1+\mu)] + 2\textrm{V'}(\mu)^2[\mu(1+\mu)] - 2\textrm{V}(\mu)\textrm{V'}(\mu)[1+2\mu]   }{\textrm{V}(\mu)^3} \right\} \),
%\(\mu^\circ_i =  \left\{\left[2\pdv{\varphi}{\mu} + \pdv[2]{\varphi}{\mu} \mu \right] \Psi^{(0)}(\mu(1+\varphi)) +  \left[(1+\varphi) + \pdv{\varphi}{\mu} \mu \right]^2 \Psi^{(1)}(\mu(1+\varphi)) \right\}
%+\left[\pdv[2]{\varphi}{\mu} \Psi^{(0)}(\varphi+2) + \left(\pdv{\varphi}{\mu}\right)^2 \Psi^{(1)}(\varphi+2) \right] -  \left\{\left[2\pdv{\varphi}{\mu} + \pdv[2]{\varphi}{\mu} (1+\mu)\right]\Psi^{(0)}(1 + (1 + \mu)(1+\varphi))+  \left[(1+\varphi) + \pdv{\varphi}{\mu} (1+\mu)\right]^2\, \Psi^{(1)}(1 + (1 + \mu)(1+\varphi))\right\}\) and \(y_i^\circ =  \left[2\pdv{\varphi}{\mu} + \pdv[2]{\varphi}{\mu} \mu \right]\log(y) - \left[2\pdv{\varphi}{\mu} + \pdv[2]{\varphi}{\mu} (1+\mu)\right]\log(1+y)  \).
Finally, we can group the values obtained above in matrix form as $\mathbf{X}^{\top} \mathbf{D}_3 \mathbf{X}$, where $\mathbf{D}_3 = \left[ c_i \, \delta_{ij}\right]$.

\vspace{0.5cm}
\noindent
\textbf{(ii)} to $\beta_j$ and $\nu_r$, for $j = 1, \ldots,p$ and $r=1,\ldots,q$, is
\begin{equation}\label{m}
\pdv[2]{\ell(\bm{\beta},\bm \nu)}{\beta_j}{\nu_r}=\sum_{i=1}^{n} \underbrace{\ddot{d}^{[i]}_{\mu\phi}\,  \,a_i\, b_i}_{m_i} \, x_{ij}\, z_{ir},
\end{equation}
where
\begin{eqnarray*}\label{deriv5}
\ddot{d}^{[i]}_{\mu\phi} = \pdv[2]{\ell(\mu_i, \phi_i)}{\mu_i}{\phi_i} &=&  \log\left(\frac{y_i}{1+y_i}\right) + \Psi^{(0)}(\mu_i(1+ \phi_i) + \phi_i +2) - \Psi^{(0)}(\mu_i(1+ \phi_i))\\
&&+ (1+\phi_i)\Psi^{(1)}(\mu_i(1+ \phi_i) + \phi_i +2)\\
&&+ \mu_i(1+\phi_i)\left[\Psi^{(1)}(\mu_i(1+ \phi_i) + \phi_i +2) - \Psi^{(1)}(\mu_i(1+ \phi_i))\right].
\end{eqnarray*}
%
%$ \ddot{d}^{[i]}_{\mu\phi} = y^{\bullet}_i - \mu^{\bullet}_i$, with $ y^{\bullet}_i =  \left[\pdv{\varphi}{\phi}  + \pdv[2]{\varphi}{\mu}{\phi} \mu \right]\log(y) - \left[\pdv{\varphi}{\phi}  %+ \pdv[2]{\varphi}{\mu}{\phi} (1+\mu) \right]\log(1+y)$,
%$\mu^{\bullet}_i =  \left\{\left[\pdv{\varphi}{\phi}  + \pdv[2]{\varphi}{\mu}{\phi} \mu \right] \Psi^{(0)}(\mu(1+\varphi))  +  \left[(1+\varphi) + \pdv{\varphi}{\phi}\mu \right] \pdv{\varphi}{\phi} \mu \Psi^{(1)}(\mu(1+\varphi)) \right\} +  \left\{ \pdv[2]{\varphi}{\mu}{\phi}  \Psi^{(0)}(\varphi+2) +  \pdv{\varphi}{\mu}\pdv{\varphi}{\phi} \Psi^{(1)}(\varphi+2)  \right \} -  \left\{\left[\pdv{\varphi}{\phi}  + \pdv[2]{\varphi}{\mu}{\phi} (1+\mu) \right]\Psi^{(0)}(1 + (1 + \mu)(1+\varphi))
%+\right. \left.  \left[(1+\varphi) + \pdv{\varphi}{\mu} (1+\mu)\right]\pdv{\varphi}{\phi} (1 + \mu)\Psi^{(1)}(1 + (1 + \mu)(1+\varphi))  \right\}$.
 The expression given in \eqref{m} can be represented in matrix form as $ \mathbf{X}^{\top} \mathbf{D}_5 \,\mathbf{Z} $, where  $\mathbf{D}_5 = \left[ m_i \, \delta_{ij}\right]$;

\vspace{0.5cm}
\noindent
\textbf{(iii)} to $\nu_j$ and $\nu_l$, for $j,l = 1, \ldots,q$, is
\begin{eqnarray}
\pdv[2]{\ell(\bm{\beta},\bm{\nu})}{\nu_j}{\nu_l}
&=&
\sum_{i=1}^{n}\underbrace{\left\{ \ddot{d}^{[i]}_{\phi}\, b_i^2  + \dot{d}^{[i]}_{\phi}\, b'_i \, b_i \right\} }_{w_i}
z_{ij}z_{il},
 \label{d2}
\end{eqnarray}
where $b'_i = \pdv{b_i}{\phi_i}$, $\dot{d}^{[i]}_{\phi_i}$ is defined in \eqref{deriv2} and
\begin{eqnarray*}\label{deriv4}
\ddot{d}^{[i]}_{\phi} = \pdv[2]{\ell(\mu_i, \phi_i)}{\phi_i} &=& -\mu_i^2\Psi^{(1)}(\mu_i(1+\phi_i) ) + (1+\mu_i)^2\Psi^{(1)}(\mu_i(1+ \phi_i) + \phi_i +2)-\Psi^{(1)}(\phi_i +2).
\end{eqnarray*}
%$ \ddot{d}^{[i]}_\phi = \left[ \frac{ \mu_i(1+\mu_i)}{\textrm{V}(\mu_i)}\right]^2 \left[  (1+\mu_i)^2 \Psi^{(1)}(1 + (1 + \mu_i)(1+\varphi_i)) - \mu_i^2 \Psi^{(1)}(\mu_i(1+\varphi_i))  - \Psi^{(1)}(\varphi_i+2)  \right]$.
%
Therefore, the expression given in \eqref{d2} can be represented in matrix form as $\mathbf{Z}^\top\,\mathbf{D}_4\, \mathbf{Z}$, where $\mathbf{D}_4 = \left[w_i\, \delta_{ij} \right]$.

Consequently, the Hessian matrix can be expressed as
\begin{equation} \mathbf{H} =\left[
           \begin{array}{cc}
            \mathbf{X}^{\top}\, \mathbf{D}_3\, \mathbf{X} &  \mathbf{X}^{\top}\, \mathbf{D}_5\,\mathbf{Z} \\
            \mathbf{Z}^{\top}\, \mathbf{D}_5\,\mathbf{X} &\mathbf{Z}^{\top}\, \mathbf{D}_4\,\mathbf{Z} \\
           \end{array} \right].
\label{hm}
\end{equation}

%It is shown in Appendix~\ref{appB} that the hessian matrix, $\mathbf{H}$,  is given by
%%Now we present an expression for Hessian matrix (HM) that is given by
% \begin{equation}\label{hm}
% \mathbf{H} = \begin{bmatrix}
%\mathbf{X}^\top\,\mathbf{D}_3\,  \mathbf{X} &  \mathbf{X}^\top \,\mathbf{D}_5\,  \mathbf{Z} \\
% \mathbf{Z}^\top \, \mathbf{D}_5\, \mathbf{X}  & \mathbf{Z}^\top\,\mathbf{D}_4\,  \mathbf{Z}
%  \end{bmatrix},
% \end{equation}
%%
%where $\mathbf{D}_3$, $\mathbf{D}_4$ and $\mathbf{D}_5$ are defined in  Appendix B.
 Taking negative expectations of \eqref{hm} leads to the expected or Fisher
information matrix for the BP regression model and can be expressed in matrix form as
\begin{equation*}\label{eqm22}
\mathbf{I}=\left[
\begin{array}{cc}
  \mathbf{X}^{\top} \mathbf{E}_3 \mathbf{X} &   \mathbf{X}^{\top} \mathbf{E}_5 \mathbf{Z} \\
 \mathbf{Z}^{\top} \mathbf{E}_5 \mathbf{X} &  \mathbf{Z}^{\top} \mathbf{E}_4 \mathbf{Z}  \\
\end{array}
\right],
\end{equation*}
where of the $i$th element of $\mathbf{E}_3$, $\mathbf{E}_4$ and $\mathbf{E}_5$ are, respectively, given by
\[
e_{3i} = (1+\phi_i)^2\left[\Psi^{(1)}(\mu_i(1+\phi_i) ) - \Psi^{(1)}(\mu_i(1+ \phi_i) + \phi_i +2)\right]\, a_i^2,
\]
\[
e_{4i} = \left[ \mu_i^2\,\Psi^{(1)}(\mu_i(1+\phi_i) ) - (1+\mu_i)^2\,\Psi^{(1)}(\mu_i(1+ \phi_i) + \phi_i +2)+\Psi^{(1)}(\phi_i +2)\right]\, b_i^2,
\]
and
\[
e_{5i} = -(1+\phi_i)\,\left\{\Psi^{(1)}(\mu_i\,(1+ \phi_i) + \phi_i +2) + \mu_i\,\left[\Psi^{(1)}(\mu_i(1+ \phi_i) + \phi_i +2) - \Psi^{(1)}(\mu_i(1+ \phi_i))\right] \right\}\,a_i\,b_i.
\]
 We can
rewrite $\mathbf{I}$ as $\widetilde{\mathbf{X}}^{\top}\widetilde{\mathbf{W}} \widetilde{\mathbf{X}}$, where
\begin{equation*}
\widetilde{\mathbf{X}}= \left[
\begin{array}{cc}
\mathbf{X}  & \mathbf{0}  \\
\mathbf{0} &\mathbf{Z}  \\
\end{array}
\right] \quad \text{and} \quad \widetilde{\mathbf{W}}= \left[
\begin{array}{cc}
\mathbf{E}_3&\mathbf{E}_5  \\
\mathbf{E}_5& \mathbf{E}_4 \\
\end{array}
\right].
\end{equation*}

Note that the parameters $\bm{\beta}$ and $\bm{\nu}$ are not orthogonal, in contrast to what is verified in the class of generalized linear regression models~\citep{McNeld:89}.

\section{Perturbation matrices}
\label{appC}

\subsection{Perturbation of the case weights}
Let $\bm{\omega}=(\omega_1, \ldots, \omega_n)^{\top}$ be a weight
vector. In this case, the perturbed log--likelihood function is
given by $ \ell(\bm{\theta},\bm{\omega}) =\sum_{i=1}^{n} \omega_i
\ell(\mu_i, \phi_i)$, where $\ell(\mu_i, \phi_i)$ is defined in
\eqref{logm}, with $0 \leq \omega_i \leq 1$, for $i = 1,\ldots,n$,
and $\bm{\omega}_0=\bm{1}^{\top}$(all-one vector). Hence, the perturbation
matrix is given by
\begin{equation*}
\widehat{\mathbf{\Delta}} = \left[\begin{array}{c}
\mathbf{X}^{\top}\, \widehat{\mathbf{D}}_1\, \widehat{\mathbf{D}}_6   \\
\mathbf{Z}^{\top}\, \widehat{\mathbf{D}}_2\, \widehat{\mathbf{D}}_7
\end{array}
\right],
\end{equation*}
where $ \mathbf{D}_6= \left[ \dot{d}^{[i]}_\mu   \, \delta_{ij}\right]$ and $ \mathbf{D}_7= \left[ \dot{d}^{[i]}_\phi   \, \delta_{ij}\right]$ .

\subsection{Perturbation of the response}
Consider now an additive perturbation on the $i$th response by
making $y_i(\omega_i) =  y_i +\omega_i s(y_i)$, where
$s(y_i)=\sqrt{\widehat{\mu}_i(1+\widehat{\mu}_i)/\widehat{\phi}_i}$ and $\omega_i \in
\mathbb{R}$, for $i=1,\ldots,n$. Then, under the scheme of response
perturbation, the log--likelihood function is given by
$\ell(\bm{\theta},\bm{\omega}) =\sum_{i=1}^{n} \ell(\mu_i,\phi_i) $, where
\begin{eqnarray*}\label{logmrp}
 \ell(\mu_i,\phi_i)&=& [\mu_i(1+\phi_i) - 1]\log(y_i(\omega_i)) - [1 + \mu_i(1 + \mu_i)(1+\phi_i)]\log(1+y_i(\omega_i)) -\log[ \Gamma(\mu_i(1+\phi_i))] \nonumber \\
&&- \log[\Gamma(\phi_i+2)] + \log[\Gamma(1 + (1 + \mu_i)(1+\phi_i))],
\end{eqnarray*}
with $\mu_i = g_1^{-1}(\mathbf{x}_i^\top\,\bm{\beta})$, $\phi_i = g_2^{-1}(\mathbf{z}_i^\top\,\bm{\nu})$ and $\bm{\omega}_0=\bm{0}$ (null vector). Hence, the perturbation matrix
here takes the form
\begin{equation*}
\widehat{\mathbf{\Delta}} = \left[\begin{array}{c}
\mathbf{X}^{\top}\, \widehat{\mathbf{D}}_1\,\widehat{\mathbf{D}}_8\,\mathbf{S} \\
\mathbf{Z}^{\top}\, \widehat{\mathbf{D}}_2\,\widehat{\mathbf{D}}_9\, \mathbf{S}
\end{array}
\right],
\end{equation*}
where $\mathbf{S} = [s(y_i) \delta_{ij}]$, the $i$th element of the diagonal matrix $\mathbf{D}_8$ is
\[
\ddot{d}_{\mu\omega}^{[i]} = (1+\phi_i) \times \frac{1}{y_i(1+y_i)},
\]
and the $i$th element of the diagonal matrix $\mathbf{D}_9$ is
\[
\ddot{d}_{\phi\omega}^{[i]} = \left[ \mu_i\times\frac{1}{y_i(1+y_i)} - \frac{1}{(1+y_i)}\right].
\]
\subsection{Perturbation of the covariates}
\subsubsection{Mean covariate perturbation}
Consider now an additive perturbation on a particular continuous
covariate, namely $x_{it}(\omega_i) = x_{it} +\omega_i\,s_i$, where $s_i$ is a scale
factor.
Then, under the scheme of regressor perturbation, the
log--likelihood function is given by $\ell(\bm{\theta},\bm{\omega})
=\sum_{i=1}^{n} \ell(\mu_i(\omega_i),\phi_i) $, where
\begin{eqnarray*}
 \ell(\mu_i(\omega_i),\phi_i)
&=& [\mu_i(\omega_i)(1+\phi_i) - 1]\log(y_i) - [1 + \mu_i(\omega_i)(1 + \mu_i(\omega_i))(1+\phi_i)]\log(1+y_i)  \nonumber \\
&&-\log[\Gamma(\mu_i(\omega_i)(1+\phi_i))]- \log[\Gamma(\phi_i+2)] + \log[\Gamma(1 + (1 + \mu_i(\omega_i))(1+\phi_i))],
\end{eqnarray*}
and $\bm{\omega}_0=\bm{0}$, with $\mu_i(\omega_i)=g_1^{-1}(\mathbf{x}_i^\top(\omega_i)\bm \beta)$ and $\mathbf x_i^\top(\omega_i)=(1,x_{i2},\ldots,x_{it}(\omega_i),\ldots,x_{ip})^\top$. Hence, the perturbation
matrix assumes the form
\[
\widehat{\mathbf{\Delta}} =
\begin{bmatrix}
s_i\,\widehat\beta_{t}\,\mathbf{X}^\top\, \widehat{\mathbf{D}}_3 + s_i\, \widehat{\mathbf{T}}_\mu\\
s_i\,\widehat\beta_{t}\,\mathbf{Z}^\top \, \widehat{\mathbf{D}}_5 \\
\end{bmatrix},
\]
%where $\mathbf{\Delta}_1$ is a
%$p\times n $ matrix with elements given by
%$$
%\mathbf{\Delta}^1_{ij}= s_x\,\widehat\beta_{t}\left[ \widehat{a}'_i\,
% \dot{d}_\mu^{[i]}\,x_{ij}  + \widehat{a}_i^2\, \ddot{d}^{[i]}_\mu\,x_{ij} \right]
% + s_x
% \left\{\begin{array}{ll}
% 0, & \text{if} \quad j\neq t; \\
% \widehat{a}_i\, \dot{d}_\mu^{[i]}, & \text{if} \quad j = t;
% \end{array}
%\right.,
%$$
%
%\[
%\Hat{\mathbf{\Delta}_1} = s_x\,\widehat\beta_{t}\,\mathbf{X}^\top\, \Hat{\mathbf{D}_1}\,\left[  \Hat{\mathbf{D}_5} +  \Hat{\mathbf{D}_1}\, \Hat{\mathbf{D}_7} \right] + s_x\, \mathbf{T}_\mu
%\]
%
where $\mathbf{T}_\mu$ is a matrix $p \times n$ given by
\[
\mathbf{T}_\mu =
\begin{blockarray}{cccccc}
1 & 2 & \cdots &n-1 & n\\
\begin{block}{[ccccc]c}
0 & 0 & \cdots &0 & 0 &1\\
0 & 0 & \cdots &0 & 0 &2\\
\vdots&\vdots&\vdots &\vdots & \vdots&\vdots \\
a_1\, \dot{d}_\mu^{[1]}&a_2\, \dot{d}_\mu^{[2]}& \cdots&a_{n-1}\, \dot{d}_\mu^{[n-1]}&a_n\, \dot{d}_\mu^{[n]} &t\\
\vdots&\vdots& \vdots &\vdots & \vdots & \vdots\\
0 & 0 & \cdots &0 & 0 & p-1\\
0 & 0 & \cdots &0 & 0 &p\\
\end{block}
\end{blockarray}.
\]
%
%and the $i$th elements of the vectors $\bm{v}_3$ and $\bm{v}_5$ are, respectively, $a_i$ and $\ddot{d}^{[i]}_{\mu\phi}$ (defined in Appendix~\ref{appA} and \ref{appB}).

\subsubsection{Precision covariate perturbation
}

Now, we additively perturb a continuous variable $\mathbf{Z}$ substituting $z_{ik}$ by $z_{ik}(\omega_i) = z_{ik} +\omega_i\,\dot{s}_i$, where $\dot{s}_i$ is a scale factor, and $\omega_i \in \mathbb{R}, \, \forall i=1,\ldots,n$. Therefore, under this scheme the log-likelihood function can be written as
\begin{eqnarray*}
\ell(\bm{\theta},\bm{\omega})
&=&\sum_{i=1}^{n} \ell(\mu_i,\phi_i(\omega_i)),\\
&=&\sum_{i=1}^{n}\left\{ [\mu_i(1+\phi_i(\omega_i)) - 1]\log(y_i) - [1 + \mu_i(1 + \mu_i)(1+\phi_i(\omega_i))]\log(1+y_i)\right.  \nonumber \\
&&\left. -\log[\Gamma(\mu_i(1+\phi_i(\omega_i)))]- \log[\Gamma(\phi_i(\omega_i)+2)] + \log[\Gamma(1 + (1 + \mu_i)(1+\phi_i(\omega_i)))]\right\},
\end{eqnarray*}
and $\bm{\omega}_0=\bm{0}$, with $\phi_i(\omega_i)=g_2^{-1}(\mathbf{z}_i^\top(\omega_i)\bm \nu)$ and $\mathbf{ z}_i^\top(\omega_i)=(1,z_{i2},\ldots,z_{it}(\omega_i),\ldots,z_{iq})^\top$. Hence, the perturbation
matrix assumes the form

\[
\widehat{\mathbf{\Delta}} =
\begin{bmatrix}
\dot{s}_i\,\widehat\nu_{k}\,\mathbf{Z}^\top\, \widehat{\mathbf{D}}_4 + \dot{s}_i\, \widehat{\mathbf{T}}_\phi\\
\dot{s}_i\,\widehat\nu_{k}\,\mathbf{X}^\top \, \widehat{\mathbf{D}}_5 \\
\end{bmatrix},
\]
where
\[
\mathbf{T}_\phi =
\begin{blockarray}{cccccc}
1 & 2 & \cdots &n-1 & n\\
\begin{block}{[ccccc]c}
0 & 0 & \cdots &0 & 0 &1\\
0 & 0 & \cdots &0 & 0 &2\\
\vdots&\vdots&\vdots &\vdots & \vdots&\vdots \\
b_1\, \dot{d}_\phi^{[1]}&b_2\, \dot{d}_\phi^{[2]}& \cdots&b_{n-1}\, \dot{d}_\phi^{[n-1]}&b_n\, \dot{d}_\phi^{[n]} &k\\
\vdots&\vdots& \vdots &\vdots & \vdots & \vdots\\
0 & 0 & \cdots &0 & 0 & q-1\\
0 & 0 & \cdots &0 & 0 &q\\
\end{block}
\end{blockarray}.
\]

\subsubsection{Simultaneous covariates perturbation}
Finally, we consider the scheme when some perturbed covariate in $\mathbf{X}$ is, also,  present in $\mathbf{Z}$. If $x_{it}=z_{ik}$, then we can define $\eta_{i2}(\omega_i) = \nu_1 + \nu_2\,z_{i2} + \cdots + \nu_k\,(x_{it}+\omega_i\,s_i) + \cdots + \nu_q\,z_{iq}$ and the non-perturbed vector is $\bm \omega_0 = \bm 0$. Thus, we can write the log-likelihood function of the following form
\begin{eqnarray*}
\ell(\bm{\theta},\bm{\omega})
&=&\sum_{i=1}^{n} \ell(\mu_i(\omega_i),\phi_i(\omega_i)),\\
&=&\sum_{i=1}^{n}\left\{ [\mu_i(\omega_i)(1+\phi_i(\omega_i)) - 1]\log(y_i) - [1 + \mu_i(\omega_i)(1 + \mu_i(\omega_i))(1+\phi_i(\omega_i))]\log(1+y_i)\right.  \nonumber \\
&&\left. -\log[\Gamma(\mu_i(\omega_i)(1+\phi_i(\omega_i)))]- \log[\Gamma(\phi_i(\omega_i)+2)] + \log[\Gamma(1 + (1 + \mu_i(\omega_i))(1+\phi_i(\omega_i)))]\right\}.
\end{eqnarray*}
Therefore, the matrix form of $\bm \Delta$ can be written as
\[
\widehat{\mathbf{\Delta}} =
\begin{bmatrix}
s_i\,\left\{\mathbf{X}^\top\,\left[\widehat\beta_{t}\, \widehat{\mathbf{D}}_3 + \widehat\nu_k\,\widehat{\mathbf{D}}_5 \right] + \widehat{\mathbf{T}}_\mu\right\} \\
\\
s_i\,\left\{\mathbf{Z}^\top\,\left[\widehat\nu_{k}\, \widehat{\mathbf{D}}_4 + \widehat\beta_t\,\widehat{\mathbf{D}}_5 \right] + \widehat{\mathbf{T}}_\phi\right\} \\
\end{bmatrix}.
\]

\bibliographystyle{authordate1}

\end{document}